\documentclass[aps,amsmath,amssymb,reprint,superscriptaddress,prd,nofootinbib,showkeys,showpacs]{revtex4-1}
\usepackage{graphicx, epsfig, bm, amsmath}
\usepackage{dcolumn}% Align table columns on decimal point
\usepackage{rotating}
\usepackage{relsize}
\usepackage{lmodern}
\usepackage{slantsc}

\usepackage{subfigure}

\ifx\pdfoutput\@undefined\usepackage[usenames,dvips]{color}
\else\usepackage[usenames,dvipsnames]{color}

\newcommand{\head}[2]{\multicolumn{1}{>{\centering\arraybackslash}p{#1}}{#2}}

\newcolumntype{C}[1]{>{\centering\arraybackslash}p{#1}}

\providecommand{\e}[1]{\ensuremath{\times 10^{#1}}}

\usepackage{array}

\providecommand{\Tr}{\text{Tr}}

\newcommand{\cosmomc}{CosmoMC}

\newcommand{\cltt}{$C_{l}^{TT}$}
\newcommand{\clee}{$C_{l}^{EE}$}
\newcommand{\clte}{$C_{l}^{TE}$}

\newcommand{\tcmb}{T_{\text{CMB}}}
\newcommand{\neff}{N_{\text{eff}}}

\newcommand{\fsky}{f_{\text{sky}}}

\newcommand{\lmax}{l_{\text{max}}}

\newcommand{\mC}{\bm{C}}

\newcommand{\mI}{\bm{I}}

\newcommand{\mM}{\bm{M}}

\begin{document}

%%%%%%%%%%%%%%%%%%%%%%%%%%%%%%%%%%%%%%%%%%%%%%%%%

\newcommand{\book}{\cite{2013neco.book.....L}}
\newcommand{\bigrev}{\cite{2006PhR...429..307L}}
\newcommand{\wong}{\cite{2011ARNPS..61...69W}}
\newcommand{\smallrev}{\cite{2012arXiv1212.6154L}}

\newcommand{\lcdm}{$\Lambda$CDM\,}
\newcommand{\lcdmw}{$\Lambda$wCDM}
\newcommand{\pcnt}{\%}
\newcommand{\lcdmwa}{$\Lambda$waCDM}

\newcommand{\sa}{Sov. Astron. Lett.}
\newcommand{\jpb}{J. Phys. B.}
\newcommand{\natu}{Nature (London)}
\newcommand{\aas}{Bull. Am. Astron. Soc.}
\def\aap{A\&A}
\def\physrep{Phys. Rep.}
\def\apj{ApJ}
\def\apjs{ApJS}
\def\apjl{ApJL}
\def\mnras{MNRAS}
\def\aj{AJ}
\def\nat{Nature}
\def\aaps{A\&A Supp.}
\def\pra{Phys.Rev.A}         % Physical Review A: General Physics
\def\prb{Phys.Rev.B}         % Physical Review B: Solid State
\def\prc{Phys.Rev.C}         % Physical Review C
\def\prd{Phys.Rev.D}         % Physical Review D
\def\prl{Phys.Rev.Lett}      % Physical Review Letters
\def\araa{ARA\&A}       % Annual Review of Astron and Astrophys
\def\gca{GeCoA}         % Geochimica et Cosmochimica Acta
\def\pasp{PASP}              % Publications of the ASP
\def\pasj{PASJ}              % Publications of the ASJ
\def\apss{ApSS}
\def\jcap{JCAP}
\def\sovast{Soviet Astronomy}
\def\plb{Phys. Lett. B}

%\preprint{AIP/123-QED}
\preprint{APS/123-QED}
\title[]{Constraining Neutrinos and Dark Energy with Galaxy Clustering in the Dark Energy Survey}
%\title[]{Constraining Neutrinos and Dark Energy with the Angular Clustering of Galaxies in the Dark Energy Survey}

\author{Alan Zablocki}
\affiliation{Department of Astronomy and Astrophysics, University of Chicago, 5640 South Ellis Avenue, Chicago, IL 60637, USA}
\affiliation{Kavli Institute for Cosmological Physics, University of
Chicago, 5640 South Ellis Avenue, Chicago, IL 60637, USA}

\date{\today}

\begin{abstract}

We determine the forecast errors on the absolute neutrino mass scale and the equation of state of dark energy by combining synthetic data from the Dark Energy Survey (DES) and the Cosmic Microwave Background
(CMB) Planck surveyor. We use angular clustering of galaxies for DES in 7 redshift shells up to $z\sim 1.7$ including cross-correlations between
different redshift shells. We study models with massless and massive neutrinos and three different dark energy models: \lcdm ($w=-1$), wCDM
(constant $w$), and waCDM (evolving equation of state parameter $w(a)=w_0 + w_{a}(1-a)$). We include the impact of uncertainties in modeling
galaxy bias using a constant and a redshift-evolving bias model. For the $\Lambda$CDM model we obtain an upper limit for the sum of 
neutrino masses from DES+Planck of $\Sigma m_\nu < 0.08$ eV (95\% C.L.) for a fiducial mass of $\Sigma m_\nu = 0.047$ eV, with a 1$\sigma$ error of 0.02 eV, assuming
perfect knowledge of galaxy bias. For the wCDM model the limit is $\Sigma m_\nu < 0.10 $ eV. For a wCDM model where galaxy bias evolves with
redshift, the upper limit on the sum of neutrino masses increases to 0.19 eV. DES will be able to place competitive upper limits on the sum of neutrino masses of 0.1-0.2 eV and could therefore strongly constrain the inverted mass hierarchy
of neutrinos. In a wCDM model the 1$\sigma$ error on constant $w$ is $\Delta w= 0.03$ from DES galaxy clustering and Planck. Allowing $\Sigma m_\nu$
as a free parameter increases the error on $w$ by a factor of 2, with $\Delta w=0.06$. In a waCDM model, in which the dark energy equation of
state varies with time, the errors are $\Delta w_0 = 0.2$ and $\Delta w_a = 0.42$. Including neutrinos and redshift dependent galaxy bias increases
the errors to $\Delta w_0 = 0.35$ and $\Delta w_a = 0.89$.\\

%Valid PACS numbers may be entered using the \verb+\pacs{#1}+ command.

\end{abstract}


%Use showkeys class option if keyword
\maketitle

\section{Introduction}
\label{1one}

Over the past decade, our understanding of the Universe has undergone a revolution driven by the huge influx of data from 
astrophysical observations. Observations of the luminosity distance to Type Ia supernovae
led to the discovery of an accelerating Universe \cite{1998AJ....116.1009R,1999ApJ...517..565P}, reviving the idea of a non-zero cosmological
constant, now independently confirmed by measurements from the Cosmic Microwave Background
(CMB) \cite{2013ApJS..208...19H,2013ApJ...779...86S,2013arXiv1303.5076P} and Large-Scale Structure (LSS) 
\cite{2004ApJ...606..702T,2006PhRvD..74l3507T,2011MNRAS.415.2876B,2011MNRAS.415.2892B}.
Analysis of these cosmological data sets has established a Standard Cosmological Model, known as Lambda+Cold Dark Matter 
(\lcdm). We live in a Universe where the majority of matter is not baryonic, but cold, dark and weakly interacting.

The accelerated expansion of the Universe suggests that most of the energy density of the Universe is in the form of dark energy with a large negative pressure \cite{2008ARA&A..46..385F}.
Despite the support for many of the theoretical ideas and the success of the standard \lcdm paradigm to explain current datasets, there remain a
number of unanswered questions regarding the fundamental physics of dark energy and neutrinos. In particular, the nature of dark energy and its 
equation of state parameter $w$ is still not accurately known. Cosmic acceleration may be the quantum energy of the vacuum or it may indicate a breakdown 
of General Relativity on cosmic scales. 

In the Standard Model of particle physics neutrinos are massless. However, results from solar and atmospheric neutrino experiments show that
neutrinos have non-zero mass (for a review see \cite{2008PhR...460....1G}) since they oscillate between the three eigenstates composed of the
three known neutrino types ($\nu_{e}$, $\nu_{\mu}$, $\nu_{\tau}$). Their individual masses and ordering of the three neutrinos, as well as 
their nature (whether neutrinos are Dirac or Majorana particles) is of fundamental importance. 

Recent neutrino oscillation experiments \cite{2012PhRvD..86g3012F} have measured a difference in the squared neutrino masses of 
$\lvert \Delta m^{2}_{31}\rvert (10^{-3}\text{eV}^2) =2.43 ^{+0.21}_{-0.22}$. This implies that
at least one eigenstate has a minimum mass of 0.047 eV. While neutrino experiments are sensitive to the difference between the square of the masses, cosmological
measurements are sensitive to the sum of neutrino masses but not much to their individual eigenstates \cite{2012JCAP...11..052H}. Thus this 
interplay between particle physics and cosmology can help in the measurement of masses and distinguish between the normal and the inverted neutrino 
hierarchy, if future measurements show $\Sigma m_\nu < 0.1$ eV. If future measurements show $\Sigma m_\nu < 0.1$ eV, then neutrino masses follow a normal hierarchy.

Massive neutrinos imprint distinct signatures on various cosmological datasets. Recent analysis of both CMB and LSS 
data have placed strong upper limits on the sum of neutrino masses. The combination of the WiggleZ 3D power spectrum and Baryon Acoustic Oscillations
(BAO) and Planck CMB data yields an upper limit of $\Sigma m_\nu < 0.18 $ eV (95\% C.L. Planck+WP+BAO+WiggleZ) \cite{2013arXiv1306.4153R}, while
\cite{2014arXiv1410.7244P} obtain an upper limit of $\Sigma m_\nu < 0.14 $ eV, when adding Ly$\alpha$ data to CMB and BAO data. Allowing $w \ne -1$ increases the upper limit to $\Sigma m_\nu < 0.49$ eV for the combination of the 3D power 
spectrum (SDSS DR9)+WP+Planck (with lensing included) \cite{2013PhRvD..88f3515G}. 

The most recent constraints on constant $w$ come from the combination of Planck+WP+BAO+JLA \footnote{Joint Light-curve Analysis of 720 
SDDS and SNLS supernovae data} where $w = -1.027 \pm 0.055$, and if dark energy varies with time, then the equation of state parameters are measured
as $w_{0} = -0.957 \pm 0.124 $ and $w_{a} = -0.336 \pm 0.552 $, assuming massless neutrinos \cite{2014arXiv1401.4064B}. Present and future photometric galaxy surveys, such
as the Dark Energy Survey (DES) \footnote{www.darkenergysurvey.org} and The Large Synoptic Survey Telescope (LSST), will be even more sensitive to the sum of neutrino masses and the equation of state parameter $w$.

DES is a multi-band, wide-field, photometric survey, covering a region of 5000 sq. deg. in the south Galactic cap in the optical \textit{griz}
filters and the Y band. The survey started on August 31 2013, and will run for 525 observing nights over 5 years. Reaching down to ~24th magnitude
in the optical, DES will measure dark energy and matter densities as well as the dark energy equation of state through four independent and
complementary methods: galaxy clustering (BAO), galaxy clusters, supernovae and weak gravitational lensing.

Forecast constraints on neutrino masses by combining large-scale structure measurements from DES and the CMB have previously been done using the
method of importance sampling by \cite{2010MNRAS.405..168L} in \lcdm models and using a Fisher Matrix analysis by \cite{2014JCAP...05..023F}. In 
this paper we carry out a full Markov Chain Monte Carlo (MCMC) 
analysis to obtain constraints on cosmological parameters by combining angular clustering of galaxies from DES with synthetic Planck CMB data
including polarization. We explore dark energy models with a constant equation of state parameter $w$, as well as dark energy models with
a time-variable equation of state $w(a)$. We also assess the increase in the errors on the sum of neutrino masses and dark energy equation of state parameters when including 
uncertainty in galaxy bias.

The paper is organized as follows: in Section \ref{sec:Theory} we discuss the theory behind the cosmological observables we use and their 
dependence on massive neutrinos and the dark energy equation of state parameter $w$. Section \ref{sec:Method} describes our data generation and
the likelihoods we use in our MCMC analysis. Results are presented in Section \ref{sec:Results} and we summarize our results in Section
\ref{sec:Discussion}. We address our assumptions in Section \ref{sec:Assumptions} and conclude in Section \ref{sec:Conclusion}. 

%%%%%%%%%%%%%%%%%%%%%%%%%%%%%%%%%%%%%%%%%%%%%%%%%%%%%%%%%%%%%%%%%%%%%%%%%%%%%%%%%%%%%%%%%%%%%

\section{Theory}
\label{sec:Theory}

We base our forecasts on theoretical observables, which include the CMB temperature and polarization anisotropy power spectra as well as the
angular clustering of galaxies. In the following sections we introduce the theory and computation of these observables and discuss how
the spectra are affected by massive neutrinos and dark energy.

\subsection{Dark Energy Parametrization}
The dark energy equation of state (EOS) parameter $w=\frac{P_{de}}{\rho_{de}}$, relates its pressure $P_{de}$ to its energy density $\rho_{de}$ and governs the
evolution of dark energy via
\begin{equation}
\frac{d\rho_{de}}{\rho_{de}} = -3 \frac{d a}{a} (1+w),  \label{eqn:Ppde}
\end{equation}
where $a$ is the Friedmann-Robertson-Walker (FRW) scale factor, with the solution $\rho_{de} \propto \text{exp} \big [-3 \int_{1}^{a} \frac{1+w(a')}{a'} da'\big ]$. While the cosmological constant
$\Lambda$ offers a simple explanation for the nature of dark energy with $w=-1$, $w$ may not be equal to $-1$ and may in fact be dynamical in 
nature. Such dark energy alternatives include scalar field models with $w>-1$ \cite{2008ARA&A..46..385F}, while phantom models, which cross 
the 'phantom divide' of $w=-1$, have $w<-1$ \cite{2002PhLB..545...23C,2005PhRvD..71d7301H}. In the case of dynamical dark energy, the EOS will
typically vary in time. We include models with constant dark energy EOS
parameter $w$ and parametrize the time evolution of dark energy using the Chevallier-Polarski-Linder (CPL) parametrization with $w(a)=w_0 + w_{a}(1-a)$
\cite{2001IJMPD..10..213C,2003PhRvL..90i1301L}, where $w_0$ is the value of $w$ today and $w_{a}$ is the rate of change of $w$ with respect to 
redshift z. In this paper we denote the dark energy density parameter with $\Omega_{\Lambda}$ if $w=-1$, and $\Omega_{de}$ otherwise.

\subsection{CMB Power Spectra}
The CMB temperature anisotropies form a scalar 2D field on the sky. It is convenient to analyze them by expansion in spherical harmonics
\begin{equation}
\frac{\Delta T}{T}(\theta,\phi)=\sum_{l}\sum_{m} a_{lm}Y_{lm}(\theta,\phi), \label{eqn:TCMBexpand}
\end{equation}
where $\Delta$T is the temperature variation from the mean, \emph{l} is the multipole, $Y_{lm}(\theta,\phi)$ 
is the spherical harmonic function of degree \emph{l} and order m, and $a_{lm}$ are the expansion coefficients or multipole moments. Assuming
the temperature anisotropies are drawn from a Gaussian field, the observed multipole moments $a_{lm}$ are Gaussian random variables with mean
zero, $\langle a_{lm}\rangle = 0$. We therefore cannot make predictions about a single $a_{lm}$. Instead, all the statistical information
in the measurement is contained in the variance of the observed $a_{lm}$'s, $\delta_{ll'}\delta_{mm'}C^{}_{l}=\langle a_{lm}^{*} a_{l'm'}^{}\rangle$, 
where $\delta_{ll'}$ is the Kronecker delta function. The measurement of the angular power spectrum $C_{l}$ has characteristic uncertainty due to finite beam size and a 
limit on the number of modes we observe on the sky known as cosmic variance, with the total error given by
\begin{equation}
\Delta C_l =\sqrt{\frac{2}{(2l+1)f_{\text{sky}}}} (C_l + N_l), \label{eqn:clerror}
\end{equation}
where $f_{\text{sky}}$ is the fraction of the sky covered by the experiment. The noise term $N_l$ is given by
\begin{equation}
 N_l = (\sigma \theta)^2 e^{l(l+1)\, \theta^2 / 8\text{ln}2 } \label{eqn:cmbnoise},
\end{equation}
where $\sigma$ and $\theta$ are the sensitivity and angular resolution respectively listed in Table \ref{tab:table1}.

While the angular power spectrum is not directly observable, we can form the rotationally invariant full-sky estimator between two fields
\begin{equation}
\hat{C}_l^{XX'}= \frac{1}{2l+1} \sum_{m=-l}^{m=l} a_{lm}^{*X} a_{lm}^{X'}, \label{eqn:clestimator}
\end{equation}
where $X,X'=\{T,E,B\}$ denotes the temperature, E-mode polarization and B-mode polarization spectra respectively. If parity is conserved then B has zero correlation
with T and E. The ensemble average of the estimator $\hat{C}^{XX'}_{l}$ is the true power spectrum $\langle \hat{C}^{XX'}_{l} \rangle = C^{XX'}_{l} $, and 
the estimator is unbiased. In the next section we describe the effects of massive neutrinos and the dark energy EOS parameter $w$ on the 
CMB temperature anisotropy $C_{l}^{TT}$. Massive neutrinos and dark energy affect the CMB anisotropies by altering the expansion history 
of the Universe and the growth rate of large-scale structure. 
\begin{figure}[t!]
\includegraphics[width=0.39\textwidth, angle=-90]{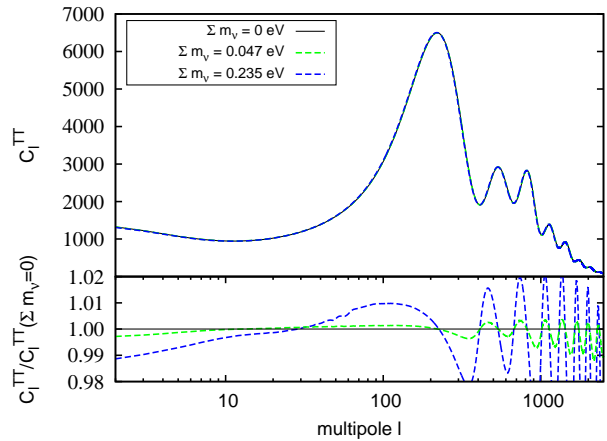}
\caption{\label{fig:figure4578} Top panel: The CMB temperature anisotropy power spectrum $C_{l}^{TT}$ for different neutrino models. The physical densities
of cold dark matter and baryons are $\omega_c= 0.1109$ and $\omega_b = 0.02258$ respectively. Shown are: a \lcdm model with massless neutrinos (black solid line), a \lcdm
model with $\sum m_{\nu} = 0.047$ eV (light green, dashed line) and a \lcdm model with $\sum m_{\nu} = 0.235$ eV (blue, dashed line). Bottom panel:
The change in $C_{l}^{TT}$ for the two neutrino models relative to the case with massless neutrinos.}
\end{figure}

\subsubsection{Massive neutrinos}

The sensitivity of the CMB to the total neutrino mass for small neutrino masses (sub-eV) is mainly due to the early Integrated Sachs-Wolfe (ISW)
effect \cite{1996ApJ...467...10D,1995PhRvD..51.2599H}. The gravitational potentials in a flat Universe without massive neutrinos, and before dark energy
starts to dominate, are constant in time. During matter domination, the density contrast $\delta =\delta \rho / \bar\rho$ grows like $a$, and 
the Poisson equation is $k^2 \psi \propto 4 \pi G a^{-1}\delta\rho / \bar\rho$, where $\psi$ is the gravitational potential in the perturbed FRW metric.
Increasing the neutrino mass, while holding baryon and cold dark matter densities fixed, increases the damping term in Eq.~(\ref{eqn:ode}) relative to the rate of growth of perturbations
since neutrinos free-stream and do not contribute to the Poisson equation (gravitational clustering). This 
imbalance leads to a decay of the gravitational potentials and results in a change in the CMB temperature at early times (early ISW) as the photons
escape gravitational potentials after decoupling, as well as at later times (late ISW) when dark energy starts to dominate.

In the top panel of Fig.~\ref{fig:figure4578} we show these effects for two neutrino density parameters, $\Omega_{\nu}=0.001$ and 
$\Omega_{\nu}=0.005$, which correspond to a sum of neutrino masses of $\sum m_{\nu} = 0.047$ eV (light green dashed line) and
$\sum m_{\nu} =0.235$ eV (blue dashed line) respectively, where $\Omega_{\nu} = \sum m_{\nu} / 93.14 h^2 \text{eV}$. We also plot a \lcdm model
with massless neutrinos (black solid line) for comparison. In the bottom panel of Fig.~\ref{fig:figure4578} we show the relative change in $C_{l}^{TT}$ for 
the two neutrino models relative to the case of massless neutrinos. Note that here, when we increase the neutrino mass, we lower the dark energy
density parameter $\Omega_{de}$ to keep a flat Universe. Hence the relative effect shown in Fig.~\ref{fig:figure4578} is not purely due to the
change in neutrino density. The physical cold dark matter and baryon densities ($\omega_c=\Omega_c h^2$ and $\omega_b = \Omega_b h^2$) remain
constant, and we keep all other parameters fixed to the fiducial cosmology of the WMAP 7th year data release \cite{2011ApJS..192...18K}.

The limited ability of CMB data to constrain neutrino masses stems from the fact that sub-eV neutrinos transition to the non-relativistic regime after photon 
decoupling ($T_{\text{dec}}\sim 0.26$ eV), and thus a single species would have to have a mass above $\sim 0.58$ eV to affect
the CMB primary anisotropies prior to recombination \cite{2009ApJS..180..330K}. Hence, the effect on the CMB for small neutrino masses is due to the background evolution
and secondary anisotropies. To get better parameter constraints, one must combine the CMB with various LSS probes, where neutrino mass effects are dominated by the inability
of neutrinos to cluster in dark matter halos due to neutrino free-streaming; the result is a suppression of the growth of structure.

Qualitatively the effects of massive neutrinos on the E-mode polarization spectrum and the cross spectrum between E-mode polarization and temperature
are similar, and improvement in parameter constraints on the sum of neutrino masses will come from breaking parameter degeneracies.

\subsubsection{Dark Energy equation of state parameter $w$}

Models with massive neutrinos and dark energy exhibit a well known parameter degeneracy
between $\sum m_{\nu}$ and $w$ in CMB and LSS data \cite{2005PhRvL..95v1301H,2008PhRvD..77f3005K}. The CMB
is sensitive to dark energy (both the equation of state parameter $w$ and the value of the density parameter $\Omega_{de}$) via the expansion rate
H(z) and the growth function D(z). Dark energy alters the amplitude of the late ISW effect \cite{1997PhRvD..55.1851C,1998PhRvL..80.1582C,1998ApJ...506..485H} 
and the angular diameter distance to the last scattering surface.

\begin{figure}[t!]
\includegraphics[width=0.39\textwidth, angle=-90]{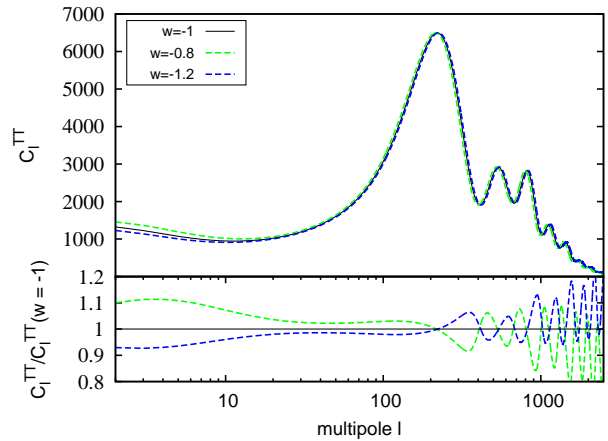}
\caption{\label{fig:figure4579} Top panel: The effect of the dark energy equation of state parameter $w$ on the CMB temperature anisotropy power
spectrum $C_{l}^{TT}$. Shown are: a \lcdm model with $w=-1.0$ (black solid line), a wCDM model with $w=-0.8$ (light green, dashed line)
and a wCDM model (blue, dashed line) with $w=-1.2$. Bottom panel: The change in $C_{l}^{TT}$ relative to the case with $w=-1$.}
\end{figure}

Increasing the value of $\Omega_{de}$ while lowering $\Omega_m$ to preserve a flat universe, reduces the rate of growth of perturbations
relative to the rate of expansion. Increasing the value of $w$
from -1 implies that dark energy starts to dominate earlier. Both effects cause the gravitational potentials to decay earlier and
lead to a greater contribution to the ISW effect. Fig. \ref{fig:figure4579} shows the increase (and decrease) in the CMB power at lower multipoles due to the late ISW effect, when $w$
is increased (decreased) from -1 while keeping other parameters fixed. The relative change in $C_{l}^{TT}$ shown in the bottom panel is a few percent at low $l$, where cosmic variance dominates. 
Another effect of dark energy is to alter the angular diameter distance to the surface of last scattering, via the integral of the inverse Hubble parameter H(z). Increasing $w$ from -1 decreases the angular diameter distance and the CMB spectrum shifts to lower 
multipoles.
\subsection{Angular Clustering of Galaxies}
\label{angcls}
% lin pk from kom5yr
In a photometric galaxy survey such as DES, we can measure the angular positions of galaxies in photometric redshift shells. The observed galaxy over-density counts $\delta^{i}_{gal}$ 
can be expressed in harmonic space, much like the decomposition of the 2D CMB temperature field. For the \textit{i}th redshift shell,
\begin{equation}
\delta^{i}_{gal}(\theta,\phi)=\sum_{l}\sum_{m} a^{i}_{lm}Y_{lm}(\theta,\phi), \label{eqn:galdenexp}
\end{equation}
with the angular auto- and cross-power spectra defined as $\delta_{ll'}\delta_{mm'}C^{gg,ij}_{l}=\langle a_{l'm'}^{i*} a_{lm}^{j}\rangle$.
Since galaxies are a biased tracer of the matter density field $\delta_{m}(z)$, to compare data to theory we need to relate the
galaxy angular power spectrum to the linear matter power spectrum P(k). The two-dimensional projection of the matter angular power spectrum 
is given by \cite{2001ApJ...555..547H,2002ApJ...571..191T}
\begin{equation}
 C_{l}= \frac{2}{\pi} \int_{0}^{\infty}  f_{l}(k)^2 P(k) k^2\,dk, \label{D1}
\end{equation}
where the functional $f_l(k)$ is the Bessel transform of a radial selection function given by
 \begin{equation}
  f_{l}(k) = \int_{0}^{\infty} j_{l}(kr)f(r)\,dr. \label{5bb}
 \end{equation}
The function $f(r)$ contains the redshift probability distribution of galaxies in the survey, $p(z)$ and the comoving distance 
$ \chi = \int_{0}^{z} \frac{c\,dz}{H(z)}$ and is equal to 
\begin{equation}
 f\left( \chi[z]=\frac{l}{k}  \right)= p(z)\left( \frac{d\chi}{dz}\right)^{-1}. \label{D4}
\end{equation}
Using the small-angle approximation and noting that at 
large $l$, the spherical Bessel function $j_l(kr)$ peaks at $l=kr$, Eq.~(\ref{5bb}) can be approximated by
 \begin{equation}
 f_{l}(k)\approx f\left(\frac{l}{k} \right) \int_{0}^{\infty} j_{l}(kr)\,dr. \label{5}
\end{equation}
Eq.~(\ref{D1}) then reduces to 
\begin{equation}
 C_{l} \approx \int_{0}^{\infty} \frac{1}{lk^2} ~\Bigg [f\left(\frac{l}{k} \right) \Bigg]^{2} P(k) k^2\,dk. \label{D5}
\end{equation}

Working in linear perturbation theory, we can relate the nearly scale-invariant spectrum of primordial fluctuations to fluctuations in the matter
density $\delta_{m}(z)$ and define the linear power spectrum P(k) as
\begin{equation}
 \frac{k^3 P(k)}{2\pi^2} = A_{s} \bigg (\frac{2k^2}{5H^2_{0} \Omega_{m} }\bigg )^2 D^2(k,z)T^2(k)  \bigg (\frac{k}{k_{0}}\bigg )^{n_{s}-1}, \label{D111}
\end{equation}
where $T(k)$ is the matter transfer function calculated with the Boltzmann code CAMB \footnote{http://camb.info/} \cite{Lewis:1999bs} and 
$D(k,z)$ is the linear growth rate. Both the matter transfer function and the linear growth rate depend on cosmological parameters.
The scalar amplitude of primordial perturbations $A_s$, and the spectral tilt $n_s$, are evaluated at a pivot scale of 
$k_{0}=0.05\text{Mpc}^{-1}$. For $n_s=1$, the spectrum is the scale-invariant Harrison-Zeldovich spectrum
\cite{1970PhRvD...1.2726H,1970A&A.....5...84Z}. Since the value of the pivot wavenumber can affect parameter constraints \cite{2007PhRvD..75h3520C},
one wants to choose a pivot scale where the estimates of $A_s$ and $n_s$ are as independent as possible 
\cite{2006PhRvD..74h3512L,2007PhRvD..75h3520C,2009ApJS..180..330K}. We have checked that for our choice of the pivot scale, $n_s$ is not strongly 
degenerate with $A_s$.

In the presence of massless neutrinos $D(k,z)$ is independent of k, but that is not the case in models with massive neutrinos. Nonetheless, we can
expand the power spectrum as $P(k,z)=P_{}(k,z=0)D(z)^{2}$ to a good approximation in linear theory. This assumption has been shown to be accurate
to better than 1\% for the redshifts and k-scales considered here, even in the presence
of massive neutrinos \cite{2006PhR...429..307L,2008PhRvD..77f3005K}. Therefore, in this work we calculate the linear matter power spectrum at $z=0$ using CAMB and solve for the linear evolution of cold dark matter perturbations governed by 
\begin{equation}
\delta^{''} + \frac{1}{a} \delta^{'} \Big( 3 + \frac{d \ln H}{d \ln a}\Big) = \frac{3}{2}
\frac{\Omega_{m} a^{-5} H_{0}^{2} \delta}{H^{2}},  \label{eqn:ode}
\end{equation}
where the prime denotes derivatives with respect to the scale factor $a$ \cite{1990eaun.book.....K}. We solve Eq.~(\ref{eqn:ode}) for the growth 
factor $\delta(a)$ and normalize it to today to get $D(a)$, with the initial conditions $D(a) = a$, and $D'(a) = 1$. We adopt the more general dark energy
parametrization with a time-variable EOS $w(a) = w_0 + w_{a}(1-a)$ with
\begin{eqnarray}
\frac{d\ln H}{d \ln a} = && -\frac{3}{2E(z)}\Big (\Omega_{m} a^{-3}+\nonumber\\
&&\Omega_{de} (1+w(a)) a^{-3(1+w_0+w_a)} e^{-3w_{a}(1-a)}\Big), \quad
% taking out omegak
%\frac{d\ln H}{d \ln a} &= -\frac{3}{2}(\Omega_{m,0} a^{-3} + \frac{2}{3} \Omega_k a^{-2} + \\
%& \Omega_{\Lambda,0} \Big( 1+w(a))a^{-3(1+w_0+w_a)} e^{-3w_{a}(1-a)}\Big)
\end{eqnarray}
where $\Omega_{m}$ and $\Omega_{de}$ are the the matter and dark energy density parameters today, and $H(z) = H_{0}E(z)$. 
In a flat Universe
\begin{align}
  E(z) = \Big(\Omega_{m}a^{-3}+\Omega_{de} a^{-3(1+w_0+w_a)} e^{-3w_{a}(1-a)} \Big)^{1/2}.  \label{eqn:EZfunc}
\end{align}
\begin{table}[t!]
\caption{\label{tab:table2}%
Best fit parameters ($\mu$ and $\sigma_z$) for the Gaussian fits to photometric redshift distributions for DES from ANNz as described in 
\cite{2008MNRAS.386.1219B,2010MNRAS.405..168L} in 7 photometric redshift shells. Also included is the fraction of galaxies in each photo-z shell.}
\begin{ruledtabular}
\begin{tabular}{c c c c}
\rule{0pt}{4ex} \parbox{1.2cm}{Photo-z shell} & \parbox{1.7cm}{Mean \\ Redshift \\ $\mu$ \smallskip}  & \parbox{1.4cm}{$\sigma_z$ \smallskip} & \parbox{2.0cm}{ Galaxy Fraction in each  shell \smallskip} \\ [0.5ex]
\colrule
\rule{0pt}{2.5ex} 0.3 $ < z < $ 0.5 & 0.405 & 0.125 & 0.211 \\
\rule{0pt}{2.5ex} 0.5 $ < z < $ 0.7 & 0.582 & 0.125 & 0.337 \\
\rule{0pt}{2.5ex} 0.7 $ < z < $ 0.9 & 0.789 & 0.123 & 0.215 \\
\rule{0pt}{2.5ex} 0.9 $ < z < $ 1.1 & 0.975 & 0.125 & 0.128 \\
\rule{0pt}{2.5ex} 1.1 $ < z < $ 1.3 & 1.203 & 0.220 & 0.098 \\
\rule{0pt}{2.5ex} 1.3 $ < z < $ 1.5 & 1.393 & 0.260 & 0.081 \\
\rule{0pt}{2.5ex} 1.5 $ < z < $ 1.7 & 1.673 & 0.291 & 0.027 \\
\end{tabular}
\end{ruledtabular}
\end{table}
Changing the integration variable in Eq.~(\ref{D5}) to $z$ and relating the galaxy over-density to the mass over-density parametrized with the
redshift dependent galaxy bias term $b(z)=\frac{\delta_{gal}(z)}{\delta_{m}(z)}$, we obtain the cross-correlation spectrum $C_{l}^{gg,ij}$
between two redshift probability distributions in the \textit{i}th and \textit{j}th shell in the Limber approximation 
\cite{1953ApJ...117..134L,1998ApJ...498...26K} as 
\begin{align}
C^{gg,ij}_l = b_{i} b_{j} \int_{0}^{\infty} dz\, P_{0} \bigg (k=\frac{l}{\chi} \bigg ) \, p^{i}(z)p^{j}(z)\frac{D(z)^{2}}{\chi(z)^{2}}\left(\frac{d\chi}{dz}\right)^{-1}, \label{kross}
\end{align}
where $P_{0}(k)$ is the matter power spectrum today. Note that in the above, we have assumed a scale-independent galaxy bias. The redshift
dependence of galaxy bias in Eq.~(\ref{kross}) is parametrized with a parameter $b_{i}$ in each photometric redshift shell.

The redshift probability distribution $p_{i}(z)$ for DES galaxies in the \textit{i}th photo-z shell, is taken from a Gaussian fit to the photometric
redshift distributions obtained using ANNz, an Artificial Neural Network code applied to simulated DES data
\cite{2008MNRAS.386.1219B,2004PASP..116..345C}. Table~\ref{tab:table2} gives the Gaussian fit parameters to the redshift distributions for DES
galaxies in 7 photometric redshift shells \cite{2008MNRAS.386.1219B,2010MNRAS.405..168L}.

The uncertainty on galaxy clustering $C^{gg,ij}_{l}$ is given by 
\begin{equation}
\Delta C^{gg,ij}_l =\sqrt{\frac{2}{(2l+1)\fsky^{\text{DES}}}} \Big (C^{gg,ij}_l + \delta_{ij} N_{\text{shot}}^{i} \Big),
\label{eqn:clerror2}
\end{equation}
where $N_{\text{shot}}^{i} = 4\pi \fsky^{\text{DES}} / N_{\text{gal,i}}$ is the shot noise contribution per redshift shell
calculated using the fractions in Table~\ref{tab:table2} and $\delta_{ij}$ is the Kronecker delta function. We assume that the total number of DES observed galaxies $N_{\text{gal}}$ is 200 million,
and the area of sky observed by DES is $\fsky^{\text{DES}}=0.125$ \cite{2005astro.ph.10346T}.
\begin{figure}[t!]
\includegraphics[width=0.39\textwidth, angle=-90]{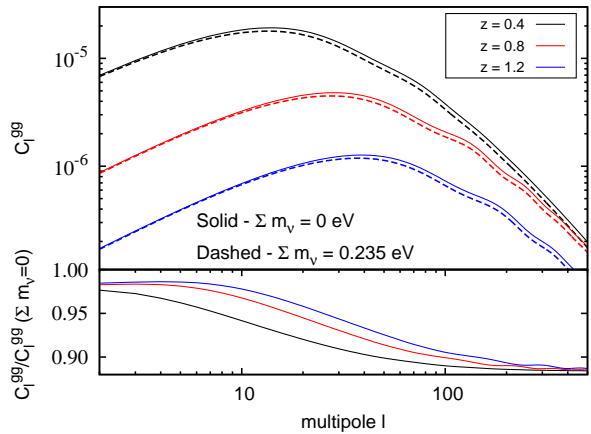}
\caption{\label{fig:figure456} Top panel: Angular clustering of galaxies in three redshift shells with mean photo-z of $z=0.4$ (black), $z=0.8$ (red) and
$z=1.2$ (blue). The solid lines depict models with massless neutrinos and dashed lines are for a \lcdm model with 
massive neutrinos, where $\Omega_{\nu}=0.005$ and $\sum m_{\nu} = 0.235$ eV. Bottom panel: The spectrum suppression relative to the case with
massless neutrinos.}
\end{figure}

\subsubsection{Massive neutrinos}
Neutrinos free-stream out of gravitational potential wells because of their large thermal velocity on scales with a wavenumber
$k > k_{\text{fs}}$ \cite{2012arXiv1212.6154L}. As neutrinos become non-relativistic during matter domination, the comoving 
free-streaming wavenumber decreases with time, and thus we expect there to be a minimum lengthscale on which neutrinos will
cluster, given by \cite{1998PhRvL..80.5255H}
\begin{equation}
k_{\text{nr}} = 0.026 \Big ( \frac{m_{\nu}}{1\text{eV}}  \Big )^{1/2} \Omega_{m}^{1/2} h \text{Mpc}^{-1}.
\end{equation}
Modes with $k < k_{\text{nr}}$ evolve like cold dark matter, and varying the neutrino mass while keeping $\Omega_m$ constant leaves the 
large-scale power spectrum invariant. In contrast, the power spectrum is suppressed by massive neutrinos on small scales, i.e., modes with 
$k >> k_{\text{nr}}$. Since on those scales neutrinos do not cluster, they can be ignored in the Poisson equation and the right hand side of Eq.~(\ref{eqn:ode})
is reduced by $(1-f_{\nu})^2$, where $f_{\nu} = \Omega_{\nu}/\Omega_m$. As a result, the growth of baryon and cold dark matter perturbations is slowed down due to the Hubble damping term in Eq.~(\ref{eqn:ode}). The overall suppression in linear theory can be fit analytically 
with $\Delta P/P \sim -8f_{\nu}$ \cite{1998PhRvL..80.5255H}. However, studies of N-body simulations that include massive neutrinos have shown that
the suppression at non-linear scales can reach $\Delta P/P \sim -10f_{\nu}$ \cite{2008JCAP...08..020B}, suggesting that the correct treatment of
non-linearities in a cosmological analysis may improve sensitivity and upper limits on the sum of neutrino masses. The above is an approximation,
and in this work we calculate the exact suppression of the linear matter power spectrum by numerically solving the Boltzmann equations using CAMB. We will study the non-linear effects on estimating the sum of neutrino masses for 
DES in future work.

In Fig. \ref{fig:figure456} we show the suppression of angular clustering due to massive neutrinos (top panel, dashed curves) in three redshift
shells relative to a model with massless neutrinos (solid curves). We keep the physical 
baryon and cold dark matter densities fixed while lowering $\Omega_{de}$ to account for a greater neutrino mass. We hold the scalar amplitude
$A_s$ constant and galaxy bias is set to $b=1$. All other parameters are fixed.
\begin{figure}[t!]
\includegraphics[width=0.39\textwidth, angle=-90]{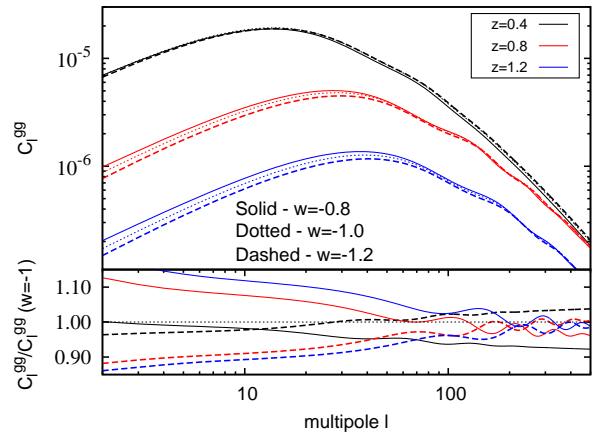}
\caption{\label{fig:figure4551} Top panel: Angular clustering of galaxies in three redshift shells with mean photo-z of $z=0.4$ (black), $z=0.8$ (red) and
$z=1.2$ (blue). The solid lines represent a model with $w=-0.8$, dotted lines show a \lcdm model with $w=-1$, whereas the dashed lines are for
a model with $w=-1.2$. Bottom panel: The relative difference in the spectra with respect to the case with $w=-1$. The scale dependence is due to transfer function T(k) effects.}
\end{figure}
\subsubsection{Dark Energy equation of state parameter $w$}

The effect of dark energy on clustering of galaxies is to alter the rate of expansion and growth of perturbations. On large scales, the change in clustering for models with different $w$
is smallest at lower redshifts, where the growth factor by definition approaches unity. At higher redshifts the difference in the rate of growth between models 
with different $w$ increases. The higher the value of $w$, the sooner dark energy starts to dominate, which causes the growth factor to asymptote 
faster, and the growth of linear perturbations to stop earlier. Since the growth factor is normalized to its value today, to match the power 
spectrum at $z=0$, the perturbations must start with a larger amplitude, hence the spectrum rises above the model with $w=-1$. In Fig. \ref{fig:figure4551}
we show angular clustering spectra in three redshift shells for three values of $w$, with galaxy bias set to $b=1$. 

%%%%%%%%%%%%%%%%%%%%%%%%%%%%%%%%%%%%%%%%%%%%%%%%%%%%%%%%%%%%%%%%%%%%%%%%%%%%%%%%%%%%%%%%%%%%%

\section{Method}
\label{sec:Method}

In this section we present the details of how we create our synthetic datasets for DES and Planck, we discuss our likelihoods, and we describe
our MCMC analysis.
\subsection{CMB Data Generation}
We generate synthetic CMB anisotropy power spectra using the Boltzmann code CAMB, computing the temperature (\cltt), the E-mode polarization (\clee) and the 
cross between the temperature and E-mode polarization (\clte) spectra up to $l=2500$. We assume white isotropic noise and Gaussian
beams, and add the noise $N_l$ (given by Eq.~\ref{eqn:cmbnoise}) to $C_l$ to create a synthetic CMB dataset. We justify this choice in Section 
\ref{sec:MCMC}. Table~\ref{tab:table1} gives the beam parameters for the 143 GHz channel used in creating the experimental noise for Planck in this work.
\begin{table}[htp]
\caption[Planck Beam Parameters]{\label{tab:table1} Planck beam parameters for the 143 GHz channel from Planck blue book \footnote{http://www.rssd.esa.int/SA/PLANCK/docs/Bluebook-ESA-SCI(2005)1\_V2.pdf} and
pre-flight performance \cite{2010A&A...520A...1T}. $\sigma_{T}$ and $\sigma_{E}$ are the sensitivity in temperature and polarization.}
\begin{ruledtabular}
\centering
\begin{tabular}{lc}
\rule{0pt}{4ex}Parameters & 143 GHz
\\ [0.5ex]
\colrule
\rule{0pt}{2.5ex}Angular Resolution $\theta$ (arcmin)          &  7.1              \\
 $\sigma_{T}$, $\Delta T / T \,(\mu K/K)$ per pixel (I)      &  2.2                    \\
 $\sigma_{E}$, $\Delta T / T \,(\mu K/K)$ per pixel (Q,U)      &  4.2            \\
\end{tabular}
\end{ruledtabular}
\end{table}
%\section{Likelihoods}
%\label{5five}
\subsection{CMB Likelihood}

Since the spherical harmonic coefficients $a_{lm}$ are Gaussian random variates and are statistically isotropic, the likelihood function is a
Wishart distribution with $\mathcal{P}(\hat{\mC_{l}}|\mC_{l}) \propto \mathcal{L}(\mC_{l}|\hat{\mC_{l}})$ and
\begin{eqnarray}
\chi_{\text{eff}}^{2} = && -2\text{ln}\mathcal{L}(\mC_{l}|\hat{\mC_{l}}) = \nonumber\\
 && \sum^{l}_{l=2} (2l+1)\left(\Tr(\hat{\mC_{l}}\mC_{l}^{-1}) + \ln\left(\frac{|\mC_{l}|}{|\hat{\mC_{l}}|}\right) - n\right), \quad
\label{eqn:cmblike}
\end{eqnarray}
where $|\mC_{l}|$ is a determinant of a covariance matrix between n correlated Gaussian fields. In
our case $n=2$ and $|\mC_{l}| = C^{TT}_{l} C^{EE}_{l}-(C^{TE}_{l})^2$ \cite{2008PhRvD..77j3013H,2009PhRvD..79h3012H}. Note that Eq.~(\ref{eqn:cmblike}) is normalized such that $\chi_{\text{eff}}^{2} = 0$, when $\hat{\mC_{l}}=\mC_{l}$. Eq.~(\ref{eqn:cmblike}) is a general case for an experiment with no noise
and full-sky coverage. In practice, experiments have noise and observe only a fraction of the sky. We modify Eq.~(\ref{eqn:cmblike}) by replacing $C_l$ with $ C_l + N_l$,
and decrease the number of modes from $(2l+1)$ to $(2l+1)\fsky$. For the CMB likelihood we assume a partial sky coverage with $\fsky = 0.85$. We do not include B-mode
polarization and ignore correlations between different multipoles due to incomplete sky coverage. We use only the 143 GHz channel and further
make the assumption that foreground removal will be optimal based on data from other channels.
\subsection{DES Data Generation}
 \begin{figure}[t!]
 \includegraphics[width=0.35\textwidth, angle=-90]{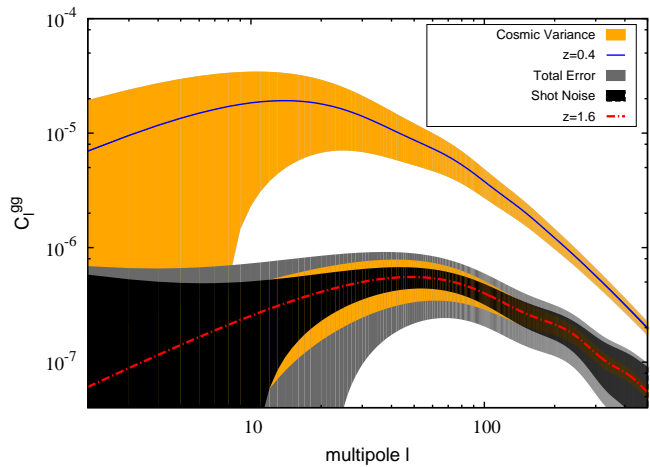}
 \caption{\label{fig:figure45c4} Angular power spectrum of galaxies with cosmic variance (orange) and shot noise (black) contributions 
 at $z=0.4$ (solid blue) and $z=1.6$ (dot dashed red), for a \lcdm model with massless neutrinos, $\Omega_{\nu}=0$. The error due to shot noise in 
 the shell at $z=0.4$ is negligible compared to cosmic variance. Galaxy bias is $b=1$.}
 \end{figure}
Our synthetic DES dataset is formed in a similar manner to the CMB dataset, where we calculate the linear power spectrum today $P_{0}(k)$ using CAMB,
and use Eq.~(\ref{kross}) to obtain the angular clustering spectra for 7 photo-z shells. The matter power spectrum is computed for 
$k_{\text{min}}$= 1\e{-4} and $k_{\text{max}}$=2.0 with a pivot point of $k=0.05 \text{Mpc}^{-1}$. For CDM models with a power spectrum amplitude consistent with Planck constraints, non-linearities become important at low redshift
for wave numbers $k>k_{\text{nl}}\sim0.2 h \text{Mpc}^{-1}$. This implies that the linear-theory approximation to the angular power spectrum will break 
down for multipoles greater than $l_{\text{nl}} = k_{\text{nl}} \chi(z)$; for $z=0.4$, $\chi(z=0.4)=1545$ Mpc, and $l_{\text{nl}} \approx 309$ (at higher redshift, 
the angular multipole of non-linearity is greater). In the likelihood analysis for DES, we therefore include only galaxy clustering multipoles
$l<l_{\text{max}}=300$. In Appendix \ref{6six_bb}, we compare these results with a more conservative non-linear cutoff of $l_{\text{max}}=100$.
Fig. \ref{fig:figure45c4} shows the evolution of the clustering signal between z=0.4 and z=1.6, with total errors according to
Eq.~(\ref{eqn:clerror2}), including shot noise and cosmic variance error contributions. Cosmic variance dominates the error budget at lower
redshifts, while the highest redshift shell is shot noise dominated at both large and small scales.

\subsection{DES Likelihood}
We write our DES likelihood in matrix form by generalizing Eq.~(\ref{eqn:cmblike}) to n spectra, including cross-correlations between
photometric redshift shells \cite{2002PhRvD..66b3528B} where
\begin{equation}
\chi_{\text{eff}}^{2} =  \sum^{l_{\text{max}}}_{l=2} (2l+1) f^{\text{DES}}_{\text{sky}} \Bigg [\ln\left(\frac{|\mM_{l}|}{|\hat{\mM_{l}}|}\right) -\Tr(\mI - \hat{\mM}_{l}^{ } \mM_{l}^{-1}) \Bigg ],\;
\label{eq:wideeq}
\end{equation}
and $|\mM_{l}|$ is the determinant of matrix $\mM_{l}$. The diagonal components of the matrix $\mM_{\it l}$ contain the theoretical power spectra with added shot noise, whereas the off-diagonal
components have shot noise equal to zero and the matrix $\mM_{\it l}$ is
\begin{align}
\mM_{\it l} = \begin{pmatrix}
\rule{0pt}{3.0ex}  C_{l}^{1,1} + N^{1}_{\text{shot}} & \cdots & C_l^{1,n} \\
\rule{0pt}{3.0ex}  \vdots & \ddots & \vdots  \\
\rule{0pt}{3.0ex}  C_l^{n,1} & \cdots & C_{l}^{n,n} + N^{n}_{\text{shot}}
 \end{pmatrix}.
 \end{align}
\subsection{MCMC Parameter Estimation}
\label{sec:MCMC}
We use \cosmomc \footnote{http://cosmologist.info/cosmomc/}, the publicly available MCMC code\cite{Lewis:2002ah},
which uses the Metropolis-Hastings algorithm \cite{Metropolis53,Hastings70} to obtain a set of random samples from the
posterior probability. We run 4-8 chains, obtaining more than 100,000 samples. We discard the first 10\% of samples as burn-in to ensure the
chain is correctly sampling the posterior. We require that the Gelman-Rubin \cite{Gelman92} convergence statistic $R-1$ is below 0.01.
We explore the (joint) DES and Planck likelihoods by adding DES and Planck chi-square. We use the May 2010 version of \cosmomc, to allow
simultaneous constraints on $w$, $w_{a}$ and $\sum m_{\nu}$. We modify both CAMB and CosmoMC by adding the Parametrized Post-Friedmann (PPF) \footnote{http://camb.info/ppf/} 
module \cite{2007PhRvD..76j4043H,2008PhRvD..78h7303F}. The PPF parametrization allows multiple crossings of $w=-1$,
the so-called 'phantom divide', and simultaneous calculation of neutrino perturbations. 

We explore cosmological models using exact theory $C_l$ rather than random realizations of a fiducial
cosmology, with the noise $N_l$ added to $C_l$ \footnote{We use the \textbf{all\_l\_exact} data
format in \cosmomc}. This allows us to measure any bias in cosmological parameter inference due to parameter degeneracies. We have checked that the
errors from such an analysis are similar to those obtained when using random realizations.

We set CAMB and CosmoMC using the same settings to minimize any bias in parameter estimation, ensuring that our results
are consistent with the input cosmology. We have found that a mismatch in settings between CAMB and CosmoMC can result in incorrect parameter estimates by up to 
2$\sigma$. The importance and magnitude of these effects is also described in \cite{2012JCAP...04..027H}.
These include:
\begin{itemize}
 \item The CMB temperature is set to $\tcmb = 2.726$K with the CMB\_outputscale parameter equal to $(\tcmb)^2$\e{12} = 7.431076\e{12}.
 \item Helium fraction $Y_{\text{He}}$ is kept constant \footnote{The flag bbn\_consistency is set to false}, $Y_{\text{He}}=0.24$.
 \item CAMB and CosmoMC calculations are set to high accuracy \footnote{accuracy\_boost=2}.
 \item Assume 3 degenerate neutrinos \footnote{This choice should not affect constraints on $\sum m_{\nu}$ \cite{2012JCAP...11..052H}.} with an effective number of relativistic species $\neff$ = 3.046 \cite{2005NuPhB.729..221M}.
 \item The reionization transition width is $\Delta z_{\text{re}}=0.5$.
\end{itemize}

Table~\ref{tab:table3} lists fiducial values for our analysis and priors on various parameters. We assume a flat universe and do not include
curvature as a free parameter while simultaneously varying dark energy, as that would require the addition of more cosmological probes, and is beyond the scope of this paper. We also do not
include CMB lensing. Weak lensing affects the anisotropy power spectrum and induces a non-Gaussianity in the lensed CMB maps by deflecting photons
from an original direction $\bm{\hat n'}$ to an observed direction $\bm{\hat n}$ on the sky. Although lensing does improve parameter constraints in \lcdm and 
wCDM models that include massive neutrinos when we only include Planck synthetic data (due to a more accurate determination of
$\sigma_8$), once we add galaxy clustering from DES, there is little or no improvement over an
\begin{figure*}[htp]
\includegraphics[width=0.9\textwidth]{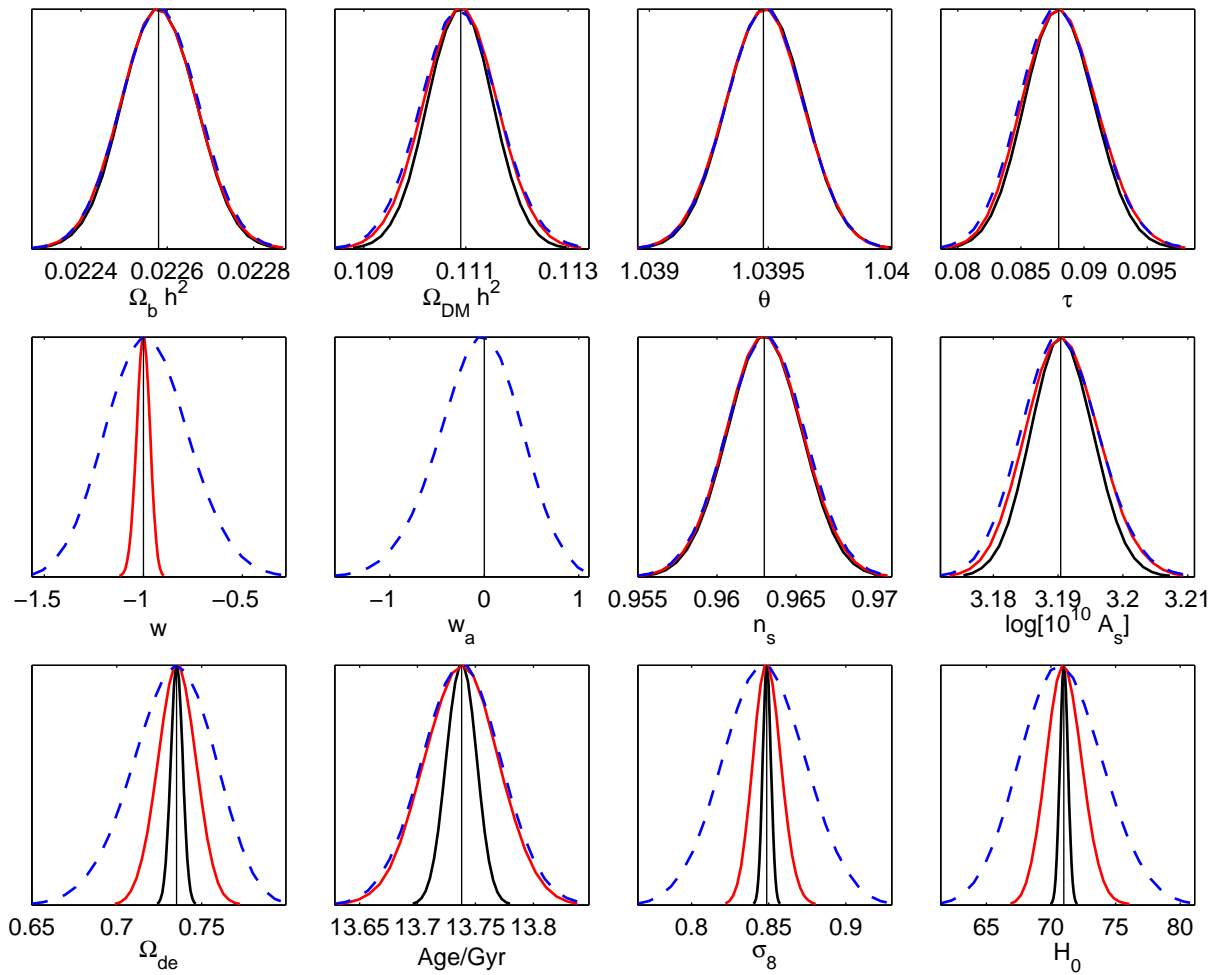}
\caption{\label{fig:fig_000wa_1D} Marginalized constraints when combining Planck and DES synthetic data for 
3 dark energy models with massless neutrinos ($\Omega_{\nu}=0$). The solid black line is the base \lcdm model, while the red line shows the wCDM model
with a constant EOS parameter $w$. The blue dashed line is a waCDM model with a time-variable equation of state in the form 
of $w(a) = w_0 + w_{a}(1-a)$. Here and throughout, $w$ stands for either $w_0$ in a waCDM model or for $w$ in a wCDM model. The y-axis is the normalized
posterior probability and the vertical bar indicates the fiducial parameter value. Galaxy bias is set to b=1.}
\end{figure*}
analysis without lensing. Furthermore, the
errors from an unlensed analysis on unlensed spectra are consistent with a lensed analysis of lensed CMB spectra, as shown in
\cite{2005PhRvD..71h3008L}.
\begin{table}[t!]
\caption{\label{tab:table3}
Fiducial values for our parameters (WMAP-7 year cosmology) and uniform priors on cosmological parameters used in our MCMC analysis.
All other priors are at their \cosmomc\, default values.}
\begin{ruledtabular}
\begin{tabular}{lccc}
\rule{0pt}{4ex}Parameters & \centering \parbox{1.2cm}{Symbol} & \centering \parbox{1.4cm}{ Fiducial \\ Value \smallskip} & Prior
\\ [0.5ex]
\colrule
\rule{0pt}{2.5ex}Baryon density &$\Omega_{b}h^{2}$        &  0.02258        &  0.005 - 0.1        \\
 Cold dark matter density      &$\Omega_{\text{DM}}h^{2}$        &  0.1109         &  0.01 - 0.99        \\
 100$r_{*}/D_A$ \footnote{100$\times$ approximation to $r_{*}/D_A$, the ratio of the sound horizon size $r_{*}$ to the angular diameter distance $D_A$ to the surface of last scattering.}             &100$\theta_{\text{MC}}$     & 1.039485        &  0.5 - 10           \\
 Spectral index                  &$n_{s}$             &  0.963          &  0.5 - 1.5        \\
 Scalar amplitude      &ln$(10^{10}A_{s})$             & 3.1904        & 2.7 - 4         \\
 Optical depth \footnote{Optical depth to Thomson scattering}  &$\tau$              &  0.088          &      0.01 - 0.8    \\
 Total neutrino mass \footnote{Masses in eV, correspond to $\Omega_{\nu} = 0.001$ and $\Omega_{\nu} = 0.005$.} &$\sum m_{\nu}$              &  0.05,0.24 & - \footnote{We use the parameter
 $f=\Omega_{\nu}/\Omega_{\text{DM}}$ with a prior of 0 - 0.3.} \\
 Dark energy EOS \footnote{Constant equation of state.}        & $w$, $w_0$             &  -1 & -2.5 - 0                \\
 Dark energy EOS \footnote{Derivative of the equation of state, as in $w(a)=w_0 + w_{a} (1-a)$.}         & $w_{a}$             &  0 & -3.5 - 1.5                \\
 Current expansion rate \footnote{Hubble constant in units of $\text{km}\text{s}^{-1}\text{Mpc}^{-1}$.}          & $H_{0}$             &  71 & 40 - 100                \\
\end{tabular}
\end{ruledtabular}
\end{table}

%%%%%%%%%%%%%%%%%%%%%%%%%%%%%%%%%%%%%%%%%%%%%%%%%%%%%%%%%%%%%%%%%%%%%%%%%%%%%%%%%%%%%%%%%%%%%

\section{Results}
\label{sec:Results}
In this section we present forecasts for how well galaxy clustering data from DES, combined with Planck, can constrain 
cosmological parameters. In Sections \ref{6six_dd} and \ref{6six_ee}, we focus on the effects of varying dark energy assumptions
in models with massless and massive neutrinos respectively. We choose two fiducial neutrino masses of $\sum m_{\nu}=0.047$ eV ($\Omega_{\nu}=0.001$) and $\sum m_{\nu}=0.235$ eV ($\Omega_{\nu}=0.005$). 
In Sections \ref{6six_hh} - \ref{6six_jj} we investigate the effects on our
constraints due to uncertainty in galaxy bias modeling. We evaluate our DES likelihood up to $l_{\text{max}}=300$ and the CMB likelihood up
to $l=2500$. In Appendix \ref{6six_bb} we show DES constraints when $l_{\text{max}}=100$ and improvement in parameter errors over a Planck only
analysis. We quote relative errors on parameters, defined as the ratio of the posterior mean to the input parameter value, and we also 
specify these errors as a fraction of the $1\sigma$ error in each model.
\subsection{Dark Energy and Neutrinos}
\label{6six_cc}
\begin{figure*}[t!]
\includegraphics[width=0.80\textwidth]{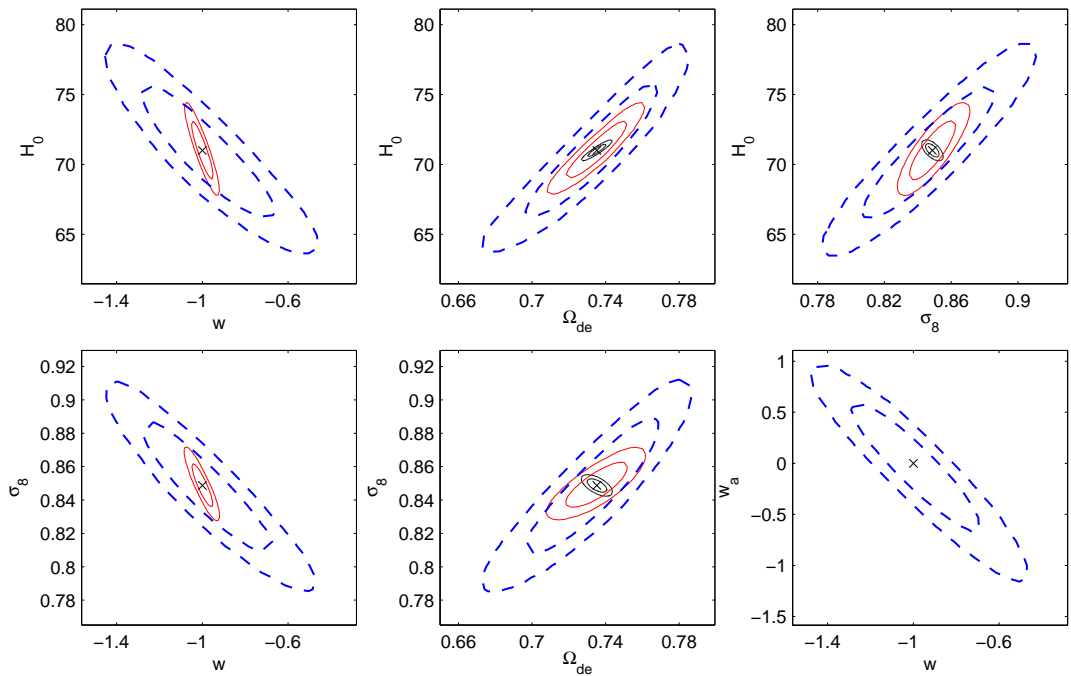}
\caption{\label{fig:fig_2D_000wa} Marginalized 68 and 95\% likelihood contours when combining Planck and DES synthetic data for
3 dark energy models with massless neutrinos ($\Omega_{\nu}=0$). The solid black line is the base \lcdm model, while the red solid line is for a 
wCDM model with a constant EOS parameter $w$. The blue dashed line is a waCDM model with a time-variable EOS $w(a)$. Galaxy bias is set to b=1. Plotting $\sigma_{8}$ vs $\Omega_{de}$ and $H_{0}$ vs $\sigma_{8}$ shows the change in the direction of degeneracies, although 
the ellipses remain centered on the fiducial parameter values marked by the symbol $\times$.}
\end{figure*}
\subsubsection{Varying Dark Energy in models with massless neutrinos}
\label{6six_dd}
We find that the constraints on the physical baryon density $\Omega_{b}h^{2}$, the ratio of sound horizon to the angular diameter distance
$r_{*}/D_A$ as well as the optical depth to reionization $\tau$, are barely affected by uncertainty in dark energy models, as these parameters are 
sensitive to physics around the decoupling epoch, before dark energy starts to dominate. They are also very well determined from CMB synthetic data
alone, and the addition of DES synthetic data does not improve constraints on these parameters considerably. 

The addition of DES data and the improvement in parameter constraints is most evident, when considering wCDM models and models with massive
neutrinos. Including galaxy clustering reduces the error on $w$ and the sum of neutrino masses $\sum m_{\nu}$ by a factor of 10 compared
to errors from Planck only, and the errors on $\sigma_8$ and $w_a$ are reduced by a factor of $\sim 2$ (see Appendix \ref{6six_bb}). 

The power spectrum amplitude (here parametrized as $ \text{log}[10^{10}A_{s}]$) and the spectral tilt $n_{s}$ errors also do not increase much when we vary the dark energy model
assumptions. When considering models with massive neutrinos and galaxy bias in later parts of this paper, we will omit constraints on 
these parameters.

Fig. \ref{fig:fig_000wa_1D} shows the marginalized posterior distributions for a cosmological model with massless neutrinos, and how parameter 
constraints are affected once we vary the dark energy model assumptions. The solid black line shows the likelihood for a base \lcdm model, with the red and blue dashed lines representing
likelihoods for wCDM with a constant EOS parameter $w$, where $w$ is free to differ from -1, and a waCDM model with time-variable 
EOS $w(a)$ respectively. We set the galaxy bias to a fiducial value of $b=1$.

Our parameter constraints in a \lcdm\,model are unbiased and the relative errors on all parameters are less than 0.1\% (0.02$\sigma$). We find that even when $w\ne-1$, we can recover the input cosmology to an accuracy of 0.1\% (0.05$\sigma$) or better,
and that parameter degeneracies are not important. In a wCDM model, we obtain $w=-1.00 \pm 0.03$.

For the dark energy model with a time-variable EOS parameter $w(a)$, the relative error in $w_0$ increases to 2.3\pcnt\,(0.11$\sigma$), while 
$w_a$ is recovered to an accuracy of 5.8\pcnt\,(0.14$\sigma$), with $w_0=-0.98 \pm 0.20$ and $w_a=-0.06 \pm 0.42$. In a waCDM model, the 
relative error on $\Omega_{de}$ is 0.4\pcnt\,(0.11$\sigma$), while the relative errors on $\sigma_{8}$ and $H_{0}$ are both 0.2\pcnt\, (0.06$\sigma$ and 0.045$\sigma$ respectively), 
signaling that parameter estimates are starting to be biased, while parameter degeneracies lead to larger error ellipses.

Some of the more degenerate parameter combinations are shown in Fig. \ref{fig:fig_2D_000wa}. In the \lcdm model, a higher Hubble constant $H_0$ 
corresponds to a lower value of $\sigma_{8}$, where a faster rate of expansion results in less structure formation and a decrease in the amplitude
of the matter power spectrum. Allowing $w\ne-1$ increases errors on cosmological parameters such as $H_0$ and $\sigma_{8}$ and results in a rotation of the $H_0$-$\sigma_{8}$
and $\sigma_{8}$-$\Omega_{de}$ degeneracies (from anti-correlated to correlated).
\begin{figure*}[htp]
 \begin{center}
 \begin{center}\includegraphics[width=0.9\textwidth]{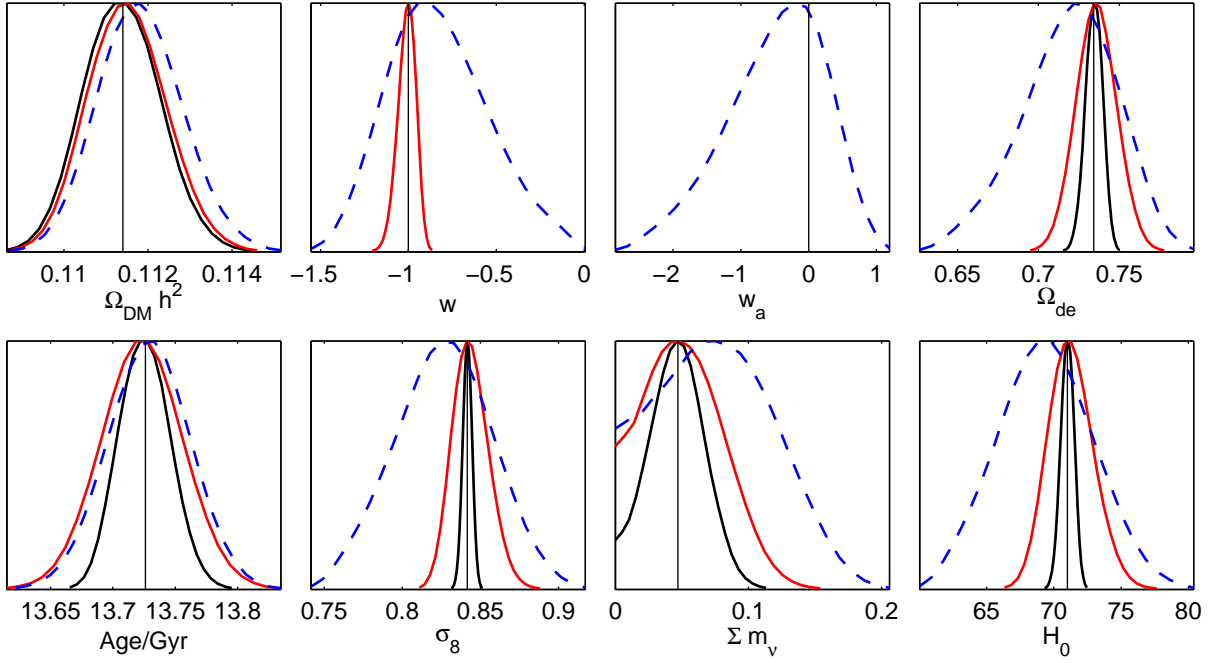}
 \caption[Short caption that will appear in list of figures]{Same as Fig. \ref{fig:fig_000wa_1D} but for a model with massive neutrinos, where $\Omega_{\nu}=0.001$ or
$\sum m_{\nu}=0.047$ eV. The parameter estimates in the waCDM$\nu$ model are highly biased, due to degeneracies between the sum of neutrino masses and other
parameters (shown in Fig. \ref{fig:fig_2D_001wa}). Estimates for the wCDM$\nu$ model (solid red line) are only slightly affected.}\label{fig:fig_001wa_1D}
 \end{center}
\end{center}\end{figure*}
\begin{figure*}[htp]
  \begin{center}
  \begin{center}\includegraphics[width=0.9\textwidth]{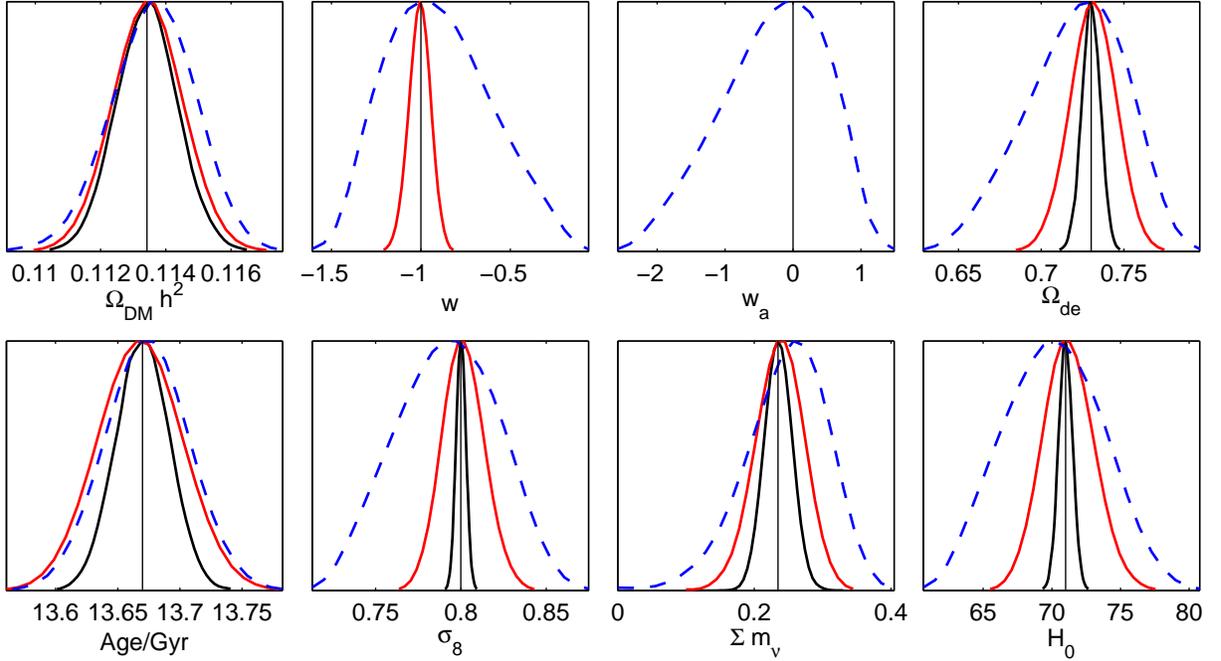}
  \caption[Short caption that will appear in list of figures]{Same as Fig. \ref{fig:fig_001wa_1D} but for a model with $\Omega_{\nu}=0.005$ or
  $\sum m_{\nu}=0.235$ eV. Since in this model we can recover the input neutrino mass, degeneracies between the sum of neutrino masses and other parameters
  are less problematic, and parameter estimates are therefore less biased.}\label{fig:fig_005wa_1D}
  \end{center}
 \end{center}\end{figure*}
 \begin{figure*}[t!]
  \begin{center}
  \begin{center}\includegraphics[width=0.80\textwidth]{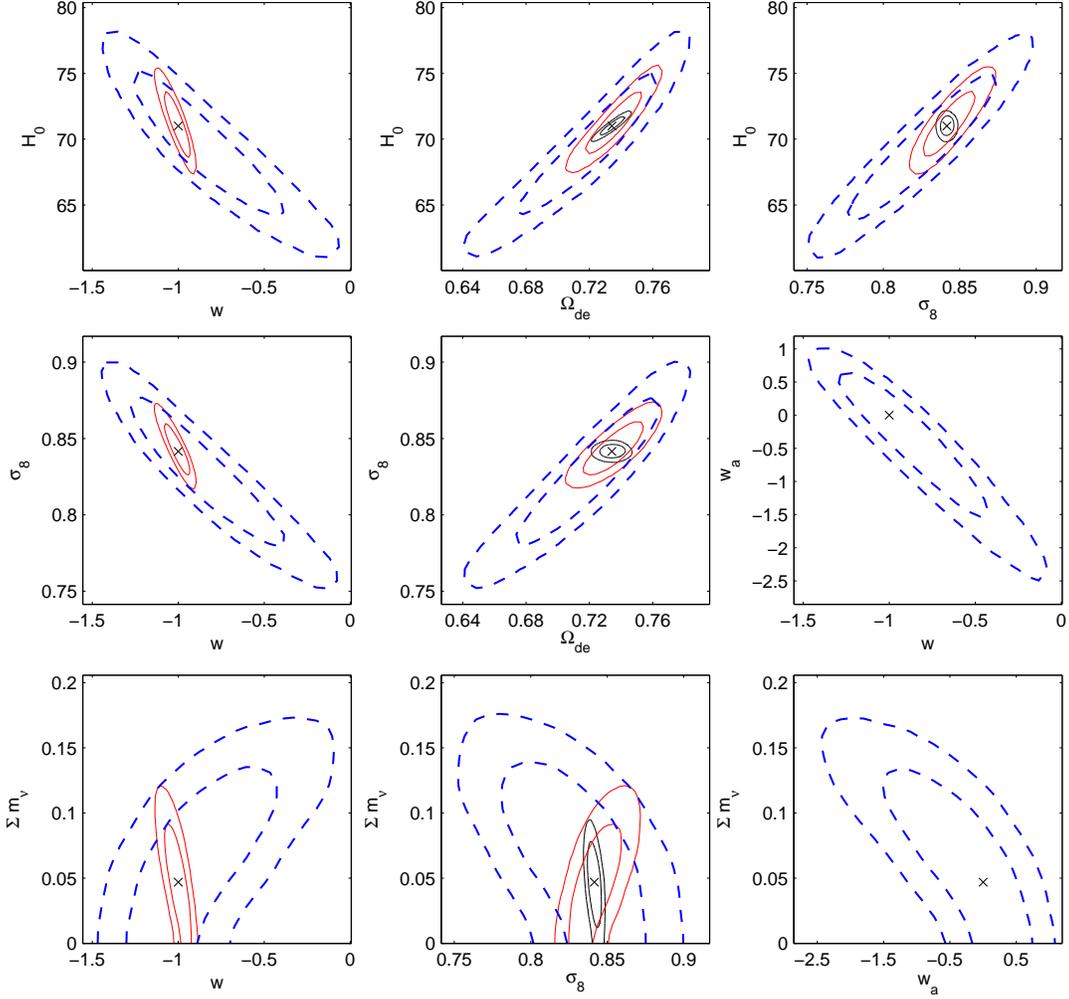}
\caption{\label{fig:fig_2D_001wa} Same as Fig. \ref{fig:fig_2D_000wa}, but for a model with massive neutrinos, 
where $\Omega_{\nu}=0.001$ or $\sum m_{\nu}=0.047$ eV. The contours in the waCDM$\nu$ model are now shifted away from the fiducial values due to
parameter degeneracies. The $\sum m_{\nu}$-$\sigma_{8}$ contours show the rotation of the degeneracies for models with massive neutrinos and constant 
$w$, relative to a model with massive neutrinos only.}
 \end{center}
 \end{center}\end{figure*}
There exists a $w$-$\sigma_{8}$ anti-correlation, due to the fact that a greater value of $w$ implies a faster rate of expansion and therefore less structure formation and a lower value of $\sigma_{8}$. The $w_0$-$w_a$ plane is poorly constrained and the degeneracy between $w_0$ and $w_a$ affects contours for all other parameters in Fig. \ref{fig:fig_2D_000wa}. 
Adding supernovae, cluster and weak lensing data from DES will improve the constraints on dark energy.
\subsubsection{Varying Dark Energy in models with massive neutrinos}
\label{6six_ee}
\begin{figure*}[t!]
\includegraphics[width=0.80\textwidth]{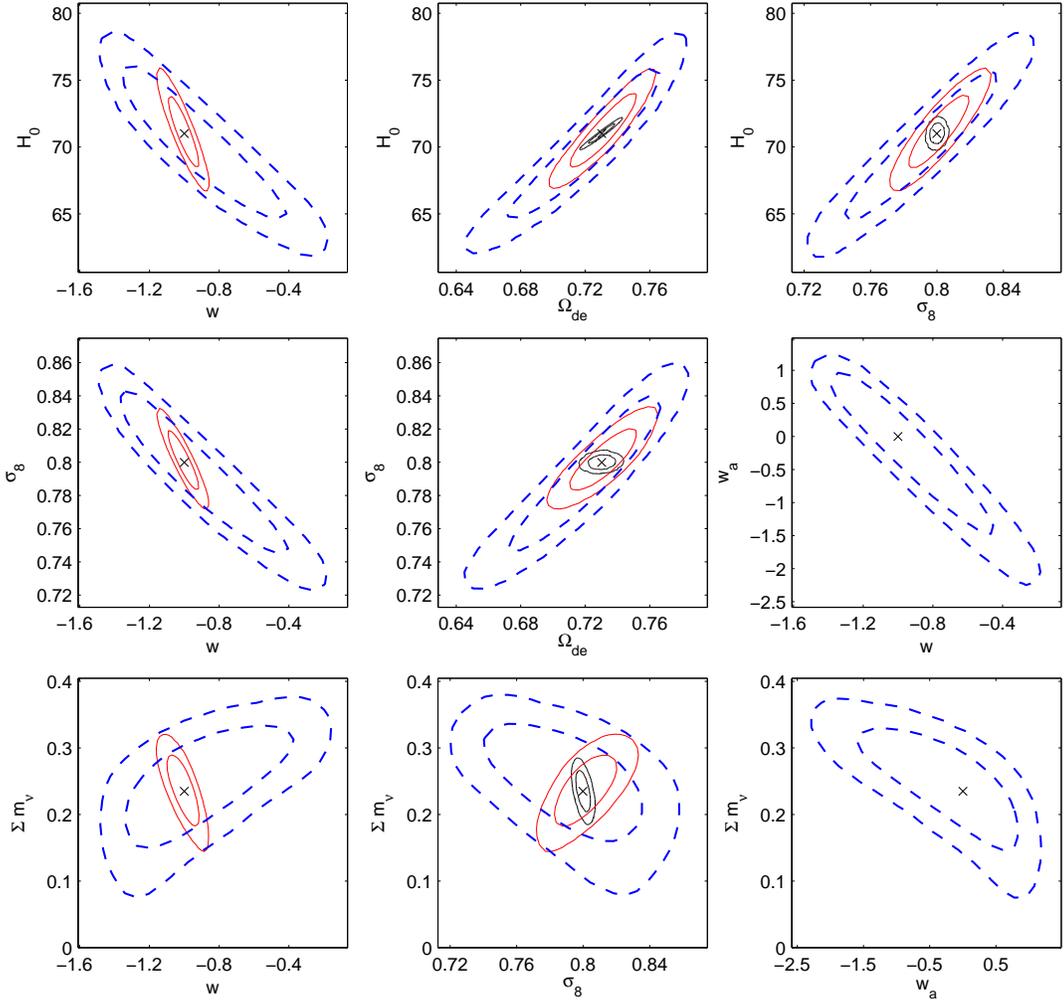}
\caption{\label{fig:fig_2D_005wa} Same as Fig. \ref{fig:fig_2D_001wa}, but for a model with massive neutrinos, where $\Omega_{\nu}=0.005$ or
$\sum m_{\nu}=0.235$ eV. Parameter degeneracies are now less severe, and so the waCDM$\nu$ contours shift more towards the fiducial parameter values, compared to contours in Fig.\ref{fig:fig_2D_001wa}.}
\end{figure*}
In this section we focus on how well we can determine the neutrino mass given an uncertainty in the equation of state parameter $w$. 
We study models where $w$ differs from -1, and when it is time-variable. We consider two models with massive neutrinos: a minimal model where 
the sum of neutrino masses is $\sum m_{\nu}= 0.047$eV (corresponding to the lower limit measured from neutrino oscillation experiments, 
and a model with a higher mass of $\sum m_{\nu}= 0.235$ eV (which is close to the upper limit from Planck \cite{2013arXiv1303.5076P}). In this section we keep the galaxy bias constant and set it to $b=1$.

In a $\Lambda$CDM$\nu$ model with massive neutrinos, the relative error on $\sum m_{\nu}$ is 2\% (0.046$\sigma$) with a measured value of $\sum m_{\nu}= 0.046 \pm 0.02$ 
and an upper limit of 0.08 eV (95\pcnt\, C.L.). Other parameters are determined to an accuracy better than 0.1\pcnt\, (0.03$\sigma$). Fig. \ref{fig:fig_001wa_1D} shows 
the marginalized posterior distributions and how parameter constraints vary for three dark energy models when
including massive neutrinos. The color scheme is as in Fig. \ref{fig:fig_000wa_1D}.

Since the prior on the fraction of neutrinos is cut off at zero, the likelihood becomes non-Gaussian with respect to $\sum m_{\nu}$.
Allowing the sum of neutrino masses to vary requires the amount of cold dark matter to increase so as to leave the matter power spectrum unchanged. This
degeneracy between $\sum m_{\nu}$ and $\Omega_{\text{DM}}h^{2}$ means that the likelihood will be also non-Gaussian with respect to $\Omega_{\text{DM}}h^{2}$. Once
we vary the dark energy equation of state, this will also increase the degeneracies in the posteriors of other parameters. 

In a wCDM$\nu$ model, there is a preference for a higher neutrino mass, and a lower value of $w$ indicating a known $w$-$\sum m_{\nu}$ degeneracy \cite{2005PhRvL..95v1301H}.
Decreasing the value of $w$ increases the amplitude in terms of $\sigma_8$, so to keep the power spectrum constant, the effect can be 
cancelled out with a higher neutrino mass. The mean value of the sum of neutrino masses is $\sum m_{\nu}= 0.0514 \pm 0.029 $ with an upper limit of 0.10 eV(95\pcnt\, C.L.), and a relative error of
10\% (0.15$\sigma$). We obtain $w=-1.01 \pm 0.05$, with a relative error of 1\% (0.2$\sigma$). The relative errors on $\sigma_{8}$ and on $H_{0}$
are 0.2\pcnt\,(0.2$\sigma$) and 0.4\pcnt\,(0.18$\sigma$) respectively. 
 
In a waCDM$\nu$ model, there is a large bias in the recovered value of the sum of neutrino masses $\sum m_{\nu}$, of around 60\%, where the 
mean of the MCMC samples is $\sum m_{\nu}= 0.0754 \pm 0.042$ with an upper limit of 0.147 eV (95\pcnt\, C.L.). This is again due to the degeneracy
between neutrino mass and $\Omega_{\text{DM}}h^{2}$ in models where we vary the EOS parameters $w_0$ and $w_a$. Fig. \ref{fig:fig_001wa_1D} shows how
the posterior distributions and parameter estimates are affected by these degeneracies. In the waCDM$\nu$ model we see a large shift in the distributions away from the 
input cosmology. Both $w_0$ and $w_{a}$ are less well constrained, with relative errors of 18\%\,(0.61$\sigma$) and 49\%\,(0.68$\sigma$) respectively, and inferred values of $w_0=-0.83 \pm 0.28$ 
and $w_a=-0.49 \pm 0.72$. Other relative errors are 0.3\pcnt\,(0.43$\sigma$) on $\Omega_{\text{DM}}h^{2}$, 2.1\% on $\Omega_{de}$(0.54$\sigma$) and 
1.8\pcnt\,(0.5$\sigma$) on $\sigma_{8}$ and 2.3\pcnt\,(0.47$\sigma$) on $H_{0}$.

Fig. \ref{fig:fig_005wa_1D} shows the likelihoods in dark energy models with a higher fiducial neutrino mass of $\sum m_{\nu}= 0.235$eV. 
The shift in the posterior distributions is less pronounced than in Fig. \ref{fig:fig_001wa_1D}, since the 
model with a higher fiducial neutrino mass disfavors lower values of $\sum m_{\nu}$, but parameter estimates are still biased. Since the lower
neutrino masses are disfavored, the inferred values of $w$ (or $w_0$) and $w_a$ are also less biased. In this case, DES+Planck combined analysis
is able to recover the neutrino mass even when we vary the dark energy EOS with time, albeit with a slightly biased estimate of the mass. 
The base $\Lambda$CDM$\nu$ model with massive neutrinos yields a measurement of $\sum m_{\nu}= 0.235 \pm 0.02$, with a 0.1\pcnt\,(0.015$\sigma$) 
relative error on $\sum m_{\nu}$. We find that our errors and upper limits on $\sum m_{\nu}$ are similar to those of \cite{2010MNRAS.405..168L} 
in \lcdm models with $w=-1$, when using DES galaxy clustering up to $l_{\text{max}} = 100$ and $l_{\text{max}} = 300$.

The relative error on $\sum m_{\nu}$ in wCDM$\nu$ is 1\pcnt\,(0.064$\sigma$) and the measured value is 
$\sum m_{\nu}= 0.237 \pm 0.034$. We obtain $w = -1.00 \pm 0.06$ with 0.7\pcnt\,(0.12$\sigma$) relative error. Relative errors on $\sigma_{8}$
and $\Omega_{de}$ are less than 0.1\pcnt(0.125$\sigma$ and 0.08$\sigma$). 

In a waCDM$\nu$ model, the relative error on $\Omega_{\text{DM}}h^{2}$ 
is 0.7\pcnt (0.16$\sigma$), while relative errors on $w_0$ and $w_{a}$ are 10\%\,(0.36$\sigma$) and 30\%\,(0.40$\sigma$) respectively, with
the mean
values of $w_0=-0.90 \pm 0.29$ and $w_a=-0.30 \pm 0.75$. Relative errors on $\Omega_{de}$, $H_{0}$ and $\sigma_{8}$ are  1.3\% (0.33$\sigma$), 1.3\% (0.25$\sigma$) 
and 1.1\% (0.28$\sigma$) respectively, while the relative error on $\sum m_{\nu}$ rises to 4.9\% (0.2$\sigma$) with the measured value being $\sum m_{\nu}= 0.246 \pm 0.06$.

The degeneracies in models with $w\ne-1$ and $w(a)$, and their improvement in cosmologies with higher neutrino masses can be understood by 
analyzing Fig. \ref{fig:fig_2D_001wa} and Fig. \ref{fig:fig_2D_005wa}, which show the marginalized 68 and 95\pcnt\, likelihood contours for
three dark energy models with the sum of neutrino masses of $\sum m_{\nu}=0.047$ eV and $\sum m_{\nu}=0.235$ eV respectively.

In a $\Lambda$CDM$\nu$ model, the higher the neutrino mass, the lower the amplitude of the power spectrum, which implies a lower value of $\sigma_8$. 
Once we allow $w$ to differ from -1, the degeneracy between $\sum m_{\nu}$ and $\sigma_8$ is rotated as seen in Fig. \ref{fig:fig_2D_001wa} (and more 
clearly in Fig. \ref{fig:fig_2D_005wa}). This degeneracy can be explained as follows:
for $w<-1$, the rate of expansion $H(z)$ is lower, compared with when $w=-1$, and this leads to more structure formation and a higher value of
$\sigma_{8}$. Decreasing $w$ increases the amplitude of the matter power spectrum, so to keep
the power spectrum constant, one has to increase the mass of the neutrinos. Hence in a wCDM$\nu$ model with massive neutrinos, the higher the neutrino
mass, the higher the value of $\sigma_8$, which is the reverse of the case in a $\Lambda$CDM$\nu$ model 
(lower middle panel of Fig. \ref{fig:fig_2D_001wa} and \ref{fig:fig_2D_005wa}).

In the waCDM$\nu$ model the likelihood contours shift along the degeneracy directions, away from the fiducial parameter values, making parameter
estimation less accurate. The change in the direction of the $w$-$\sum m_{\nu}$ degeneracy is non-trivial and is due to a much
larger $w_0$-$w_{a}$ space allowed, since those parameters are not well constrained.

As we increase $\Omega_{\text{DM}}h^{2}$ and $\sum m_{\nu}$, the 
expansion rate increases, which can be mimicked by a more positive value of $w$, and less structure formation, which lowers the value of $\sigma_8$.
A higher $\Omega_{\text{DM}}h^{2}$ now allows more positive values of $w$, whereas in a wCDM$\nu$ model with massive neutrinos the reverse was true.

\begin{figure}[htp]
\includegraphics[width=0.34\textwidth,angle=-90]{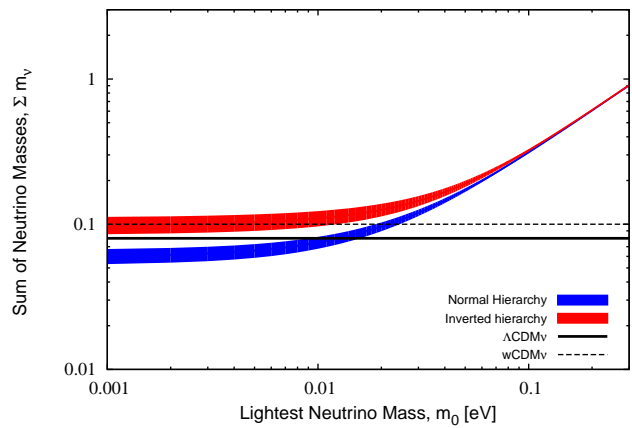}
\caption[Neutrino mass hierarchy and upper limits from DES and Planck]{The sum of neutrino masses plotted as a function of the lightest state $m_0$ for the inverted ($m_0=m_3$) and normal ($m_0=m_1$) hierarchy.
The filled contours are the 3$\sigma$ regions allowed by the results of solar and atmospheric neutrino flavor oscillations. Overplotted are the upper limits from our DES + Planck analysis, showing
that in $\Lambda$CDM ($\Sigma m_\nu < 0.08$ eV (95\% C.L.)) and wCDM ($\Sigma m_\nu < 0.10$ eV (95\% C.L.)), the combination of DES
and CMB data could distinguish between the normal and inverted hierarchy provided that we have full knowledge
of galaxy bias.\label{fig:figureNH}}
\end{figure}

To summarize: As we change the dark energy model from one with a cosmological constant to $w\ne-1$ or a time-variable equation of 
state, the upper limits on the sum of neutrino masses $\sum m_{\nu}$ are 0.08 eV, 0.10 eV and 0.147 eV (95\pcnt\, C.L.) respectively. The first two
upper limits are especially interesting since they would effectively probe the neutrino mass hierarchy, and suggest a possibility of constraining
the inverted hierarchy, if not excluding it, should it turn out that the measured $\sum m_{\nu}$ is less than 0.1 eV at high confidence and if galaxy bias is known.
In Fig. \ref{fig:figureNH} we show the sum of neutrino masses as a function of the lightest state for the normal and inverted hierarchy, highlighting the
upper limits from DES and Planck. These limits are quite competitive compared to current bounds on $\sum m_{\nu}$, even if $w\ne-1$ or $w_{a}\ne0$, despite the fact that
we have not used any supplementary data for DES from either supernovae, galaxy clusters or weak lensing.
\subsection{Effect of Galaxy Bias in models with massive neutrinos and dark energy}
\label{6six_ff}

In the previous section we showed cosmological constraints when combining synthetic data from DES with simulated Planck CMB spectra assuming perfect
knowledge of galaxy bias. In this section we consider constraints on the same models, but now include uncertainty in galaxy bias since galaxies
are biased tracers of dark matter.

We assume a linear scale-independent bias for the galaxy over-density with $\delta_{gal}(z)=b_{gal}(z)\delta_{m}(z)$ where
$\delta_{m}(z)$ is the matter over-density. Our choice is motivated by the fact that scale-dependent bias is small on 
large scales, where linear theory is thought to accurately describe the matter perturbations
\cite{2009MNRAS.392..682C,2005MNRAS.362..505C,2007ApJ...657..645P}. We consider two bias models; one where a single global bias parameter $b$ is
used at all redshifts with a fiducial value of $b=1$, and a second model, where we allow the amplitude of galaxy bias to evolve with redshift.

Instead of an explicit function of redshift, we instead choose to parametrize redshift evolution of galaxy bias with a parameter $b_i$ for each of
the redshift slices. This assumption is justified since we expect the bias evolution to be a smooth monotonic function of redshift. 
\begin{table}[htp]
\scriptsize
\caption{\label{tab:table4aaa}The values for the redshift-evolving galaxy bias model used in our analysis.}
\begin{ruledtabular}
\begin{tabular}{c c c c c c c c}
\rule{0pt}{2.5ex}z & 0.4 & 0.6 & 0.8 & 1.0 & 1.2 & 1.4 & 1.6
\\ [0.5ex]
\colrule
\rule{0pt}{3.0ex} $b_{i}$ &1.45 & 1.6&1.78 &1.97 & 2.19& 2.39&2.59\\
\end{tabular}
\end{ruledtabular}
\end{table}
When running our MCMC chains we either vary a single bias parameter $b$ (those models are denoted with $+b$) or we vary the seven bias
parameters $b_i$ (those models are denoted with $+b_i$). Table~\ref{tab:table4aaa} lists the values of galaxy bias for our fiducial 
redshift-evolving bias model. This second model is based on a fit to N-body simulations found in Fig. 1 of \cite{2012MNRAS.422.2904G}, where the
authors measure the galaxy bias from the MICE N-body simulations. We use flat priors on all bias parameters, with a typical width of 0.25, i.e.,
$\pm 0.125$. We have checked that the data constrains all bias parameters more than the priors in each of our models.
\begin{figure*}[htp]
\includegraphics[width=0.79\textwidth]{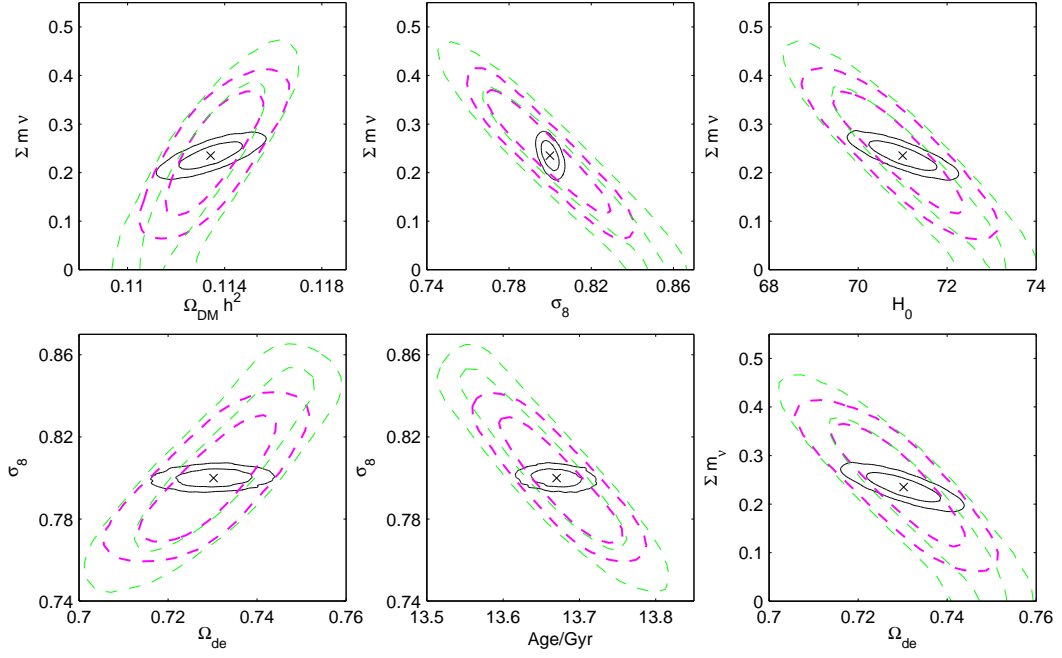}
\caption{\label{fig:fig_005bb7_2D} Marginalized 68 and 95\% likelihood contours from DES and Planck for a $\Lambda$CDM$\nu$ model with massive
neutrinos and $\Omega_{\nu}=0.005$ or $\sum m_{\nu}=0.235$ eV, taking into account uncertainty in galaxy bias modeling. The solid black line is
the base $\Lambda$CDM$\nu$ model assuming a full knowledge of galaxy bias by setting $b=1$. The green thin dashed line is
a $\Lambda$CDM$\nu+b$ model with a single free bias parameter $b$. The magenta thick
dashed line is a $\Lambda$CDM$\nu+b_i$ model with a redshift-evolving bias, parametrized with 7 free bias parameters
$b_{i}$, one per redshift bin. The fiducial cosmology is marked by the symbol $\times$.}
\end{figure*}
\begin{figure*}[htp]
\includegraphics[width=0.79\textwidth]{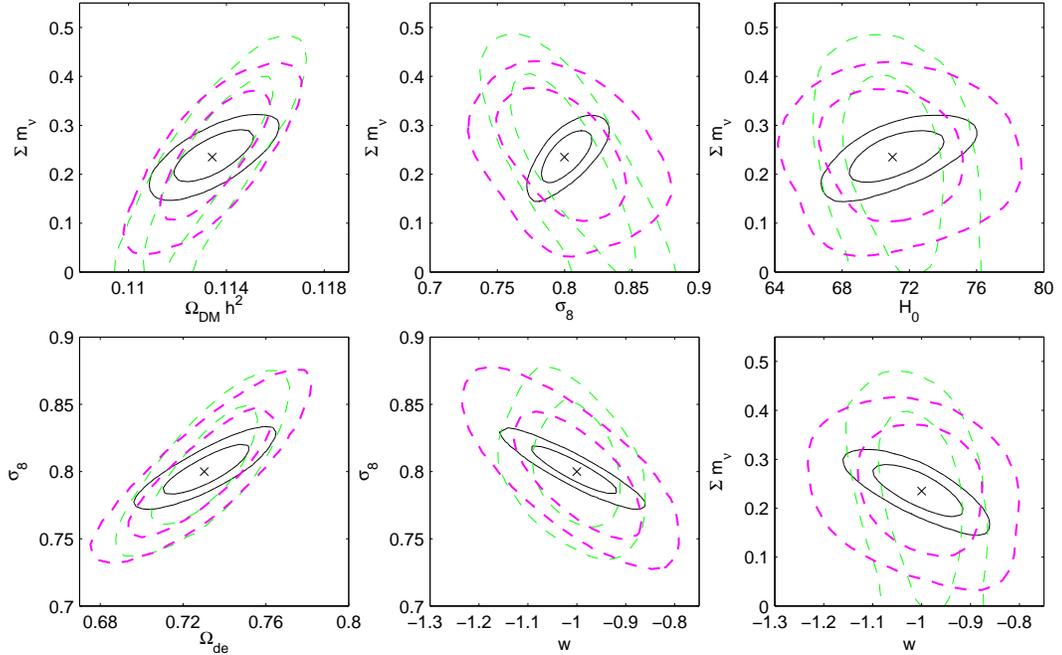}
\caption{\label{fig:fig_005wbb7_2D} 
Same as Fig. \ref{fig:fig_005bb7_2D} but for a wCDM$\nu$ model with massive neutrinos and $\Omega_{\nu}=0.005$ or $\sum m_{\nu}=0.235$ eV and
a constant EOS parameter $w$, where $w\ne-1$.}
\end{figure*}
\subsubsection{Galaxy bias effects in models with massive neutrinos only}
\label{6six_hh}

In this section we consider the effects of uncertainty in galaxy bias on parameter constraints in models with massive neutrinos.
The addition of new 'amplitude' parameters such as galaxy bias increases the overall errors on all parameters relative to the base
$\Lambda$CDM$\nu$ model and opens up the degeneracy in $\sigma_{8}$ and $\sum m_{\nu}$. We discuss uncertainties in galaxy bias as well as 
degeneracies with other parameters in Appendix \ref{bias_plots}. We expect the galaxy bias $b$ and $\sigma_{8}$ to be 
anti-correlated since the power spectrum is proportional to their product $P(k)\propto b^2 \sigma^2_{8}$. An increase in the neutrino mass results
in a smaller value of $\sigma_{8}$, and therefore the higher the galaxy bias the greater the neutrino mass.

Fig. \ref{fig:fig_005bb7_2D} shows the joint two-dimensional marginalized constraints for the base
$\Lambda$CDM$\nu$ model (solid black line) with our two bias models over-plotted; a single bias model (green dashed line) and a redshift 
evolving bias model (magenta). Taking into account the redshift evolution of galaxy bias improves the parameter errors over
the single bias model, but the parameter constraints are not as strong as when $b$ is held fixed. We find that a seven parameter bias model
allows a more than 3$\sigma$ determination of the neutrino mass, whereas a single bias model does not exclude the lower $\sum m_{\nu}$ region.
The measured value of $\sum m_{\nu}$ in a model with bias fixed was $\sum m_{\nu}= 0.235 \pm 0.02$. In contrast, a single bias model recovers the 
fiducial mass to within 9.2\pcnt\,(0.18$\sigma$), with $\sum m_{\nu}= 0.216 \pm 0.11$, whereas a model with redshift-evolving galaxy bias over-predicts
the value by 3\pcnt\,(0.1$\sigma$), with $\sum m_{\nu}= 0.243 \pm 0.076$. 

Thus both bias models introduce a shift in the recovered parameter values, though the 
shift is smaller with a seven parameter bias model. The parameters $\Omega_{\text{DM}}h^{2}$, $\Omega_\Lambda$, $H_0$ and $\sigma_{8}$ are all determined
to better than 0.6\% (0.2$\sigma$) in a single bias model, and to better than 0.2\% (0.12$\sigma$) in the seven parameter bias model.
\begin{figure*}[t!]
\includegraphics[width=0.79\textwidth]{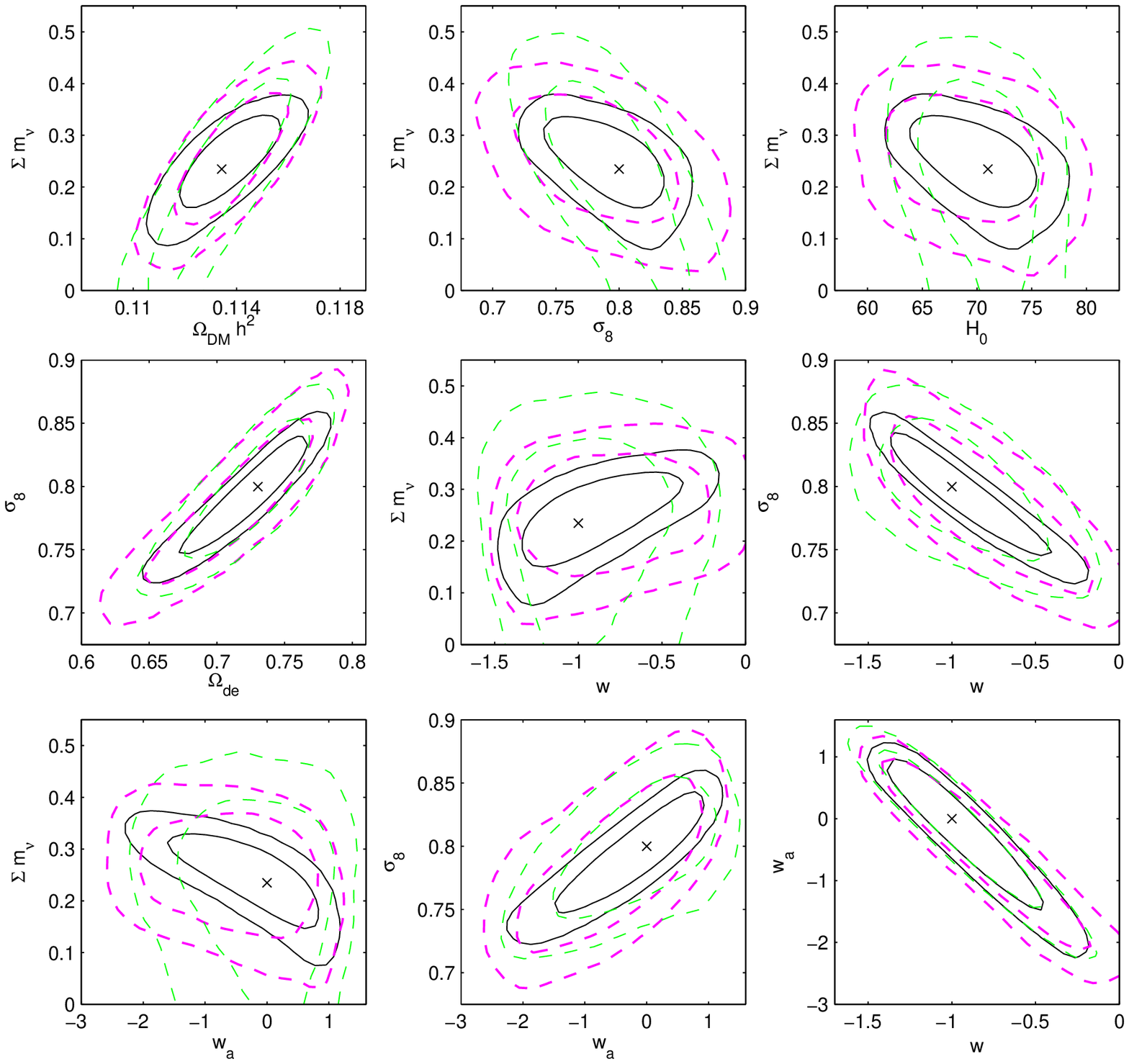}
\caption{\label{fig:fig_005wabb7_2D} Same as Fig. \ref{fig:fig_005wbb7_2D} but for a waCDM$\nu$ model with massive neutrinos and $\Omega_{\nu}=0.005$ or $\sum m_{\nu}=0.235$ eV and
a time-variable EOS parameter $w(a)$.}
\end{figure*}
\subsubsection{Galaxy bias effects in models with massive neutrinos and $w$}
\label{6six_ii}

In Fig. \ref{fig:fig_005wbb7_2D} we show the 68 and 95\% likelihood contours for models with massive neutrinos, a constant EOS parameter $w$
and our two bias models. We find that we can still recover the fiducial value of $\sum m_{\nu}$ in a wCDM$\nu$ model with massive neutrinos when we 
include seven additional bias parameters in our likelihood. The measured value of $\sum m_{\nu}$ in a model with $w\ne-1$ and fixed bias
was $\sum m_{\nu}= 0.237 \pm 0.034$. In contrast, the estimate for the neutrino mass in a wCDM$\nu$ model with a single bias parameter is
$\sum m_{\nu}= 0.226 \pm 0.11$, whereas a model with evolving redshift recovers the neutrino mass to 1.5 \% (0.04$\sigma$) with 
$\sum m_{\nu}= 0.239 \pm 0.08$. We find that once we include galaxy bias, adding $w$ as a free parameter in models with massive neutrinos 
does not increase the neutrino mass errors or the upper limits on $\sum m_{\nu}$ appreciably. 

In a model with $\sum m_{\nu}= 0.235$ eV, the error on the dark energy equation of state does not increase when we add a single bias parameter. When $b$ was fixed, the constraint was
$w = -1.00 \pm 0.06$ and when $b$ is varied we also obtain $w = -1.00 \pm 0.06$. The error does increase in a model with redshift-evolving bias 
for which there are 7 extra parameters, and the recovered value is $w = -1.00 \pm 0.088$. In Appendix \ref{bias_plots} we present errors on the
measurement of galaxy bias and we show the marginalized probability distributions along with the 2D likelihood contours for models with $w=-1$ 
and $w\ne-1$ in the presence of massive neutrinos.

\subsubsection{Galaxy bias effects in dynamical dark energy models with massive neutrinos}
\label{6six_jj}
In this section we consider the waCDM$\nu$ model with a time-variable EOS and massive neutrinos (black solid line in 
Fig. \ref{fig:fig_005wabb7_2D} ). We combine this base model with 
our two bias models. Once we allow $w_a$ to vary, the degeneracy between $w_0$ and $w_a$ increases the errors on the other parameters,
most notably on $\sum m_{\nu}$, $H_0$, $\Omega_{de}$ and $\sigma_8$. As a comparison, when we held galaxy bias fixed and varied dark energy
in a model with massive neutrinos, the recovered values were $w_0 = -0.9 \pm 0.29 $ and $w_a = -0.30 \pm 0.75$, with the sum of neutrino masses of 
$\sum m_{\nu}= 0.246 \pm 0.06$. In contrast to those results, the single bias parameter model in Fig. \ref{fig:fig_005wabb7_2D}
(green dashed line) gives $w_0 = -0.93 \pm 0.32$, $w_a = -0.22 \pm 0.82$ and $\sum m_{\nu}= 0.234 \pm 0.116$. When allowing for a 
redshift-evolving bias, the results (shown with a magenta dashed line) are $w_0 = -0.80 \pm 0.35$, $w_a = -0.57 \pm 0.88$ and 
$\sum m_{\nu}= 0.251 \pm 0.08$.
\subsubsection{Bias amplitude effects and massive neutrinos}
\label{6six_kkk}
\begin{figure*}[htp]
\begin{center}
\includegraphics[width=0.45\textwidth]{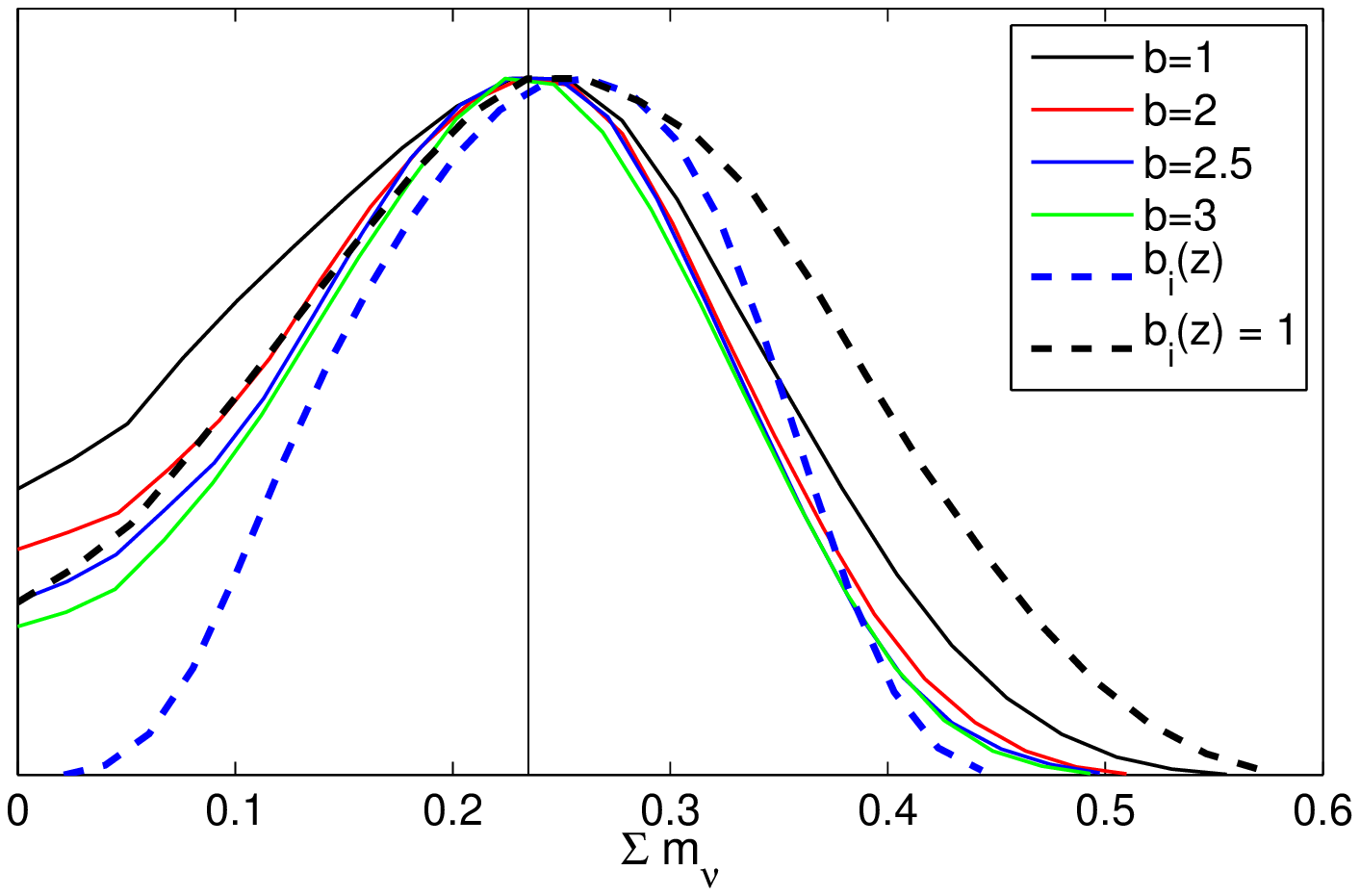}
\includegraphics[width=0.463\textwidth]{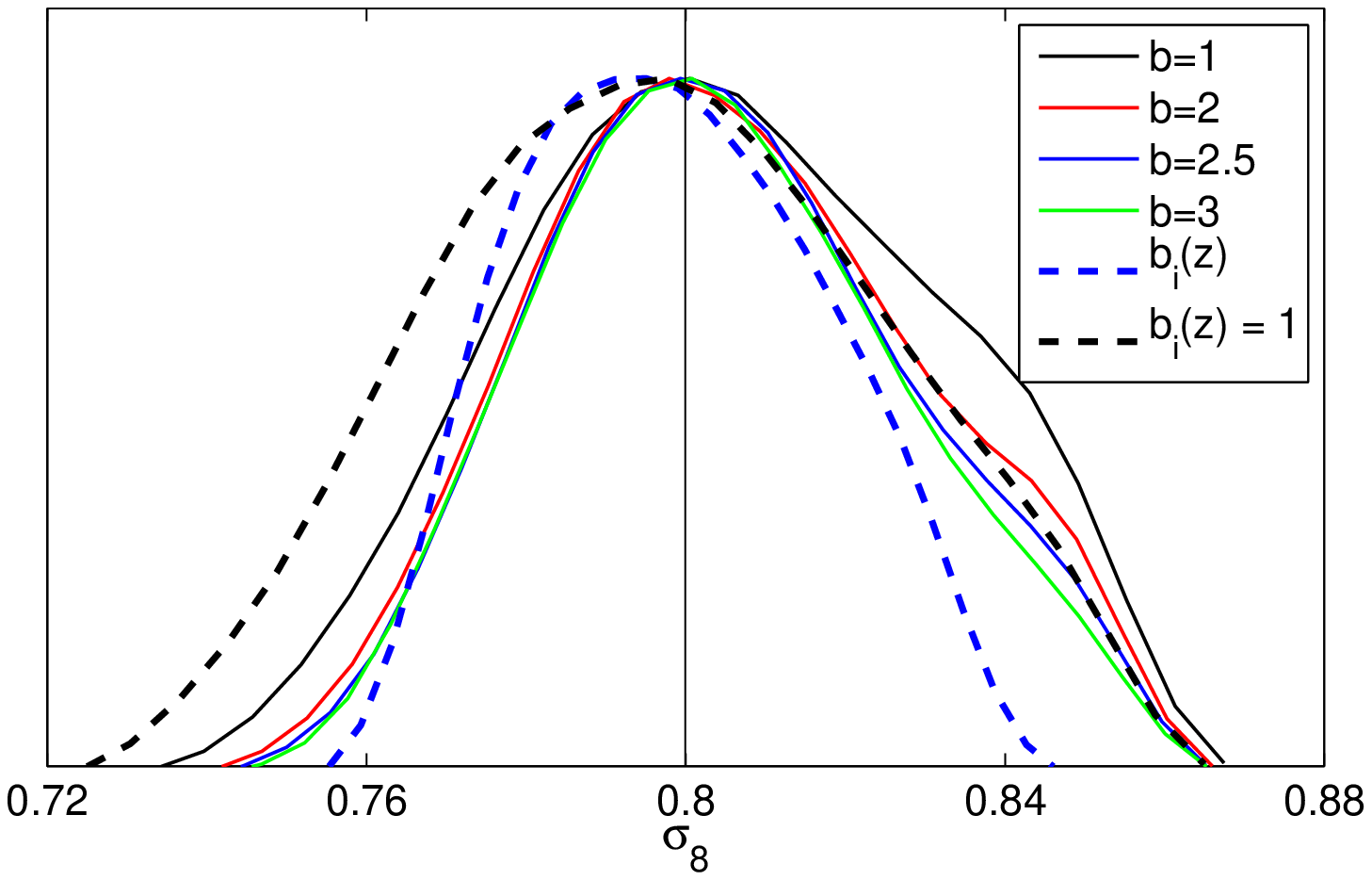}
\end{center}
\caption{Marginalized posterior distributions for $\sum m_{\nu}$ and $\sigma_8$ where $\sum m_{\nu_{}}=0.235$ eV. Shown are six models with
different assumptions about the galaxy bias amplitude. The four solid curves are for a single bias model with b = \{1,2,2.5,3.0\} and the two
dashed curves are for a seven parameter model: a redshift-evolving model with the bias values from Table~\ref{tab:table4aaa} and
a second model which has all $b_i$ set to 1.}
\label{adc:fig333}
\end{figure*}

So far we have been comparing models where $b=1$ to a model with a redshift-evolving bias which has a higher signal to noise ratio in $C^{gg}_l$. One could argue that we should have set the fiducial value of $b$ in the 
single-parameter model to its value at the mean redshift for DES (using the distributions in Table~\ref{tab:table2}, the mean redshift is 
$\bar z = 0.8$, with $b(\bar z) = 1.78$). We find that even when we use a higher value for galaxy bias ($b(\bar z) = 1.78$, and higher), our
constraints on neutrino masses are still not as strong those that use the seven parameter model. The improvement over the single bias model is
therefore only partially due to higher signal to noise. That this is the case can be seen in Fig.~\ref{adc:fig333}, where we show the likelihoods for $\sum m_{\nu}$ and $\sigma_8$ in 6 models with 
different assumptions  about the amplitude of galaxy bias. All models are include massive neutrinos, where $\sum m_{\nu} = 0.235$ eV. 
The four solid curves in black, red, blue and green are for the single bias model with $b=\{1,2,2.5,3\}$ respectively. The two dashed curves
show the seven parameter model where the fiducial values of bias are either all equal to one, or to the values in our redshift-evolving bias model. 
In the latter model, the amplitude of the power spectrum is increased in each redshift shell, which is equivalent to higher values of $\sigma_8$.
This can also be interpreted as a model with a lower fiducial neutrino mass, where the neutrinos do not suppress power on small scales as much.
For each $b_i$ we are setting an upper limit on the amplitude for the spectrum, where for a larger amplitude, one would require an even smaller
neutrino mass. Thus an upper limit on $b_i$ corresponds to excluding neutrino masses less than some minimum $\sum m_{\nu}$, which in our case 
helps in 'detection' or recovering the input value for the sum of neutrino masses. We predict that this effect will be even more important when
non-linear scales are used in obtaining neutrino mass constraints.
\begin{figure*}[htp]
\begin{center}
\includegraphics[width=0.45\textwidth]{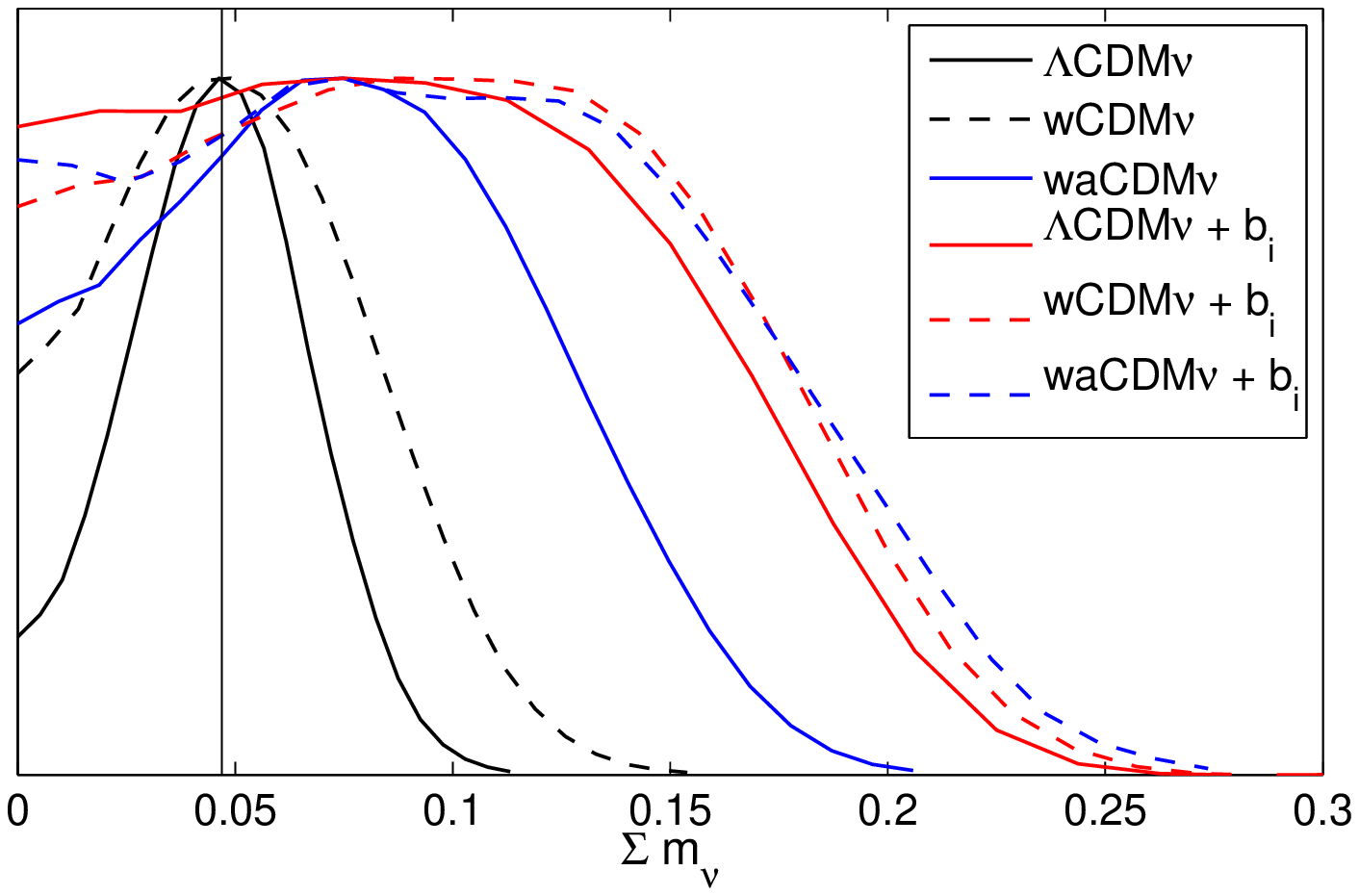}
\includegraphics[width=0.445\textwidth]{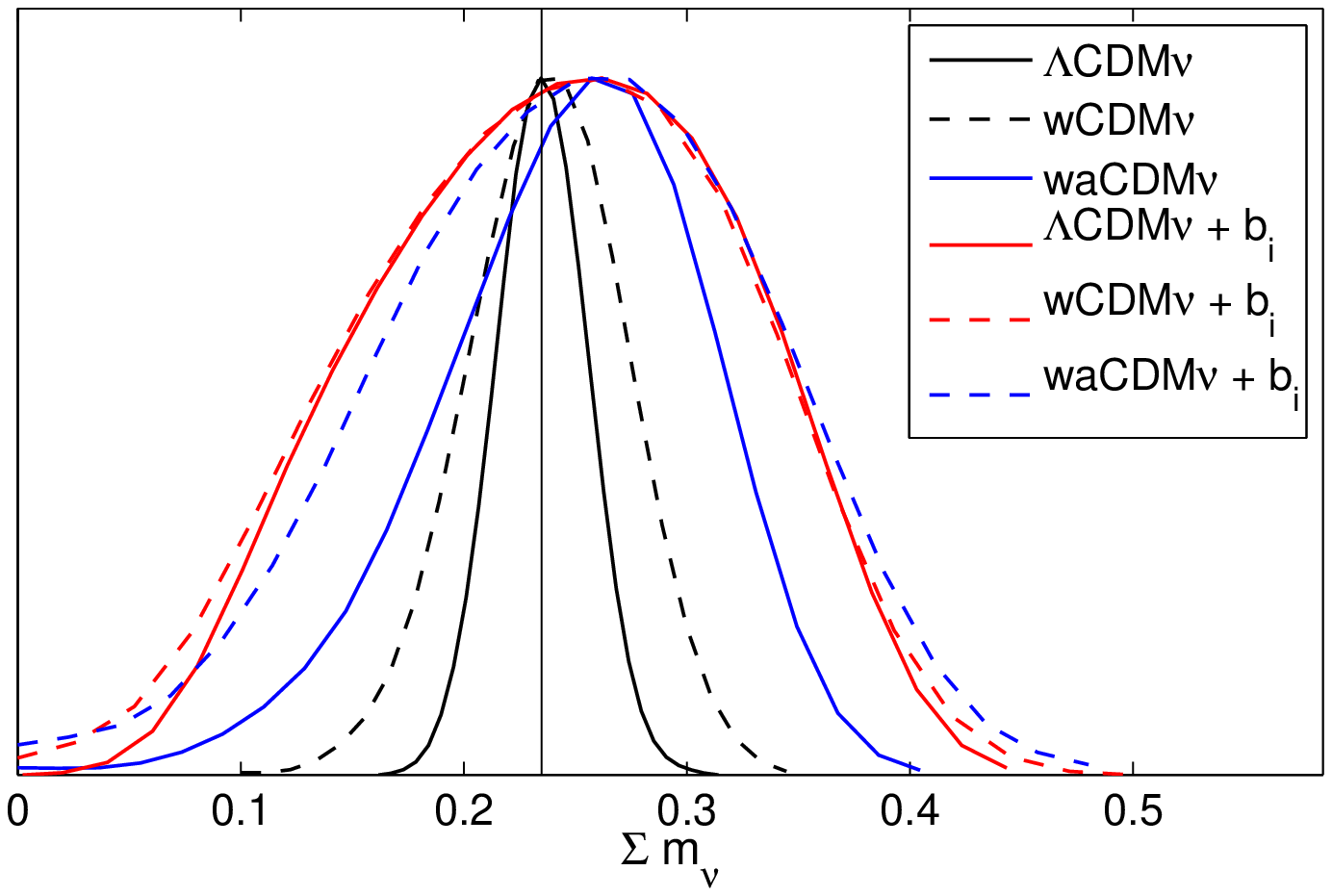}
\end{center}
\caption{\label{fig:figure10} Marginalized posterior distributions with $\sum m_{\nu_{}}=0.047$ eV (left panel) and $\sum m_{\nu_{}}=0.235$ eV (right panel).
Shown are dark energy models with $w=-1$, $w\ne-1$ and time-varying equation of state $w(a)$ for models with $b=1$ and a redshift-evolving bias model.}
\label{adc:fig3} 
\end{figure*}
\begin{figure*}[htp]
\includegraphics[width=0.45\textwidth]{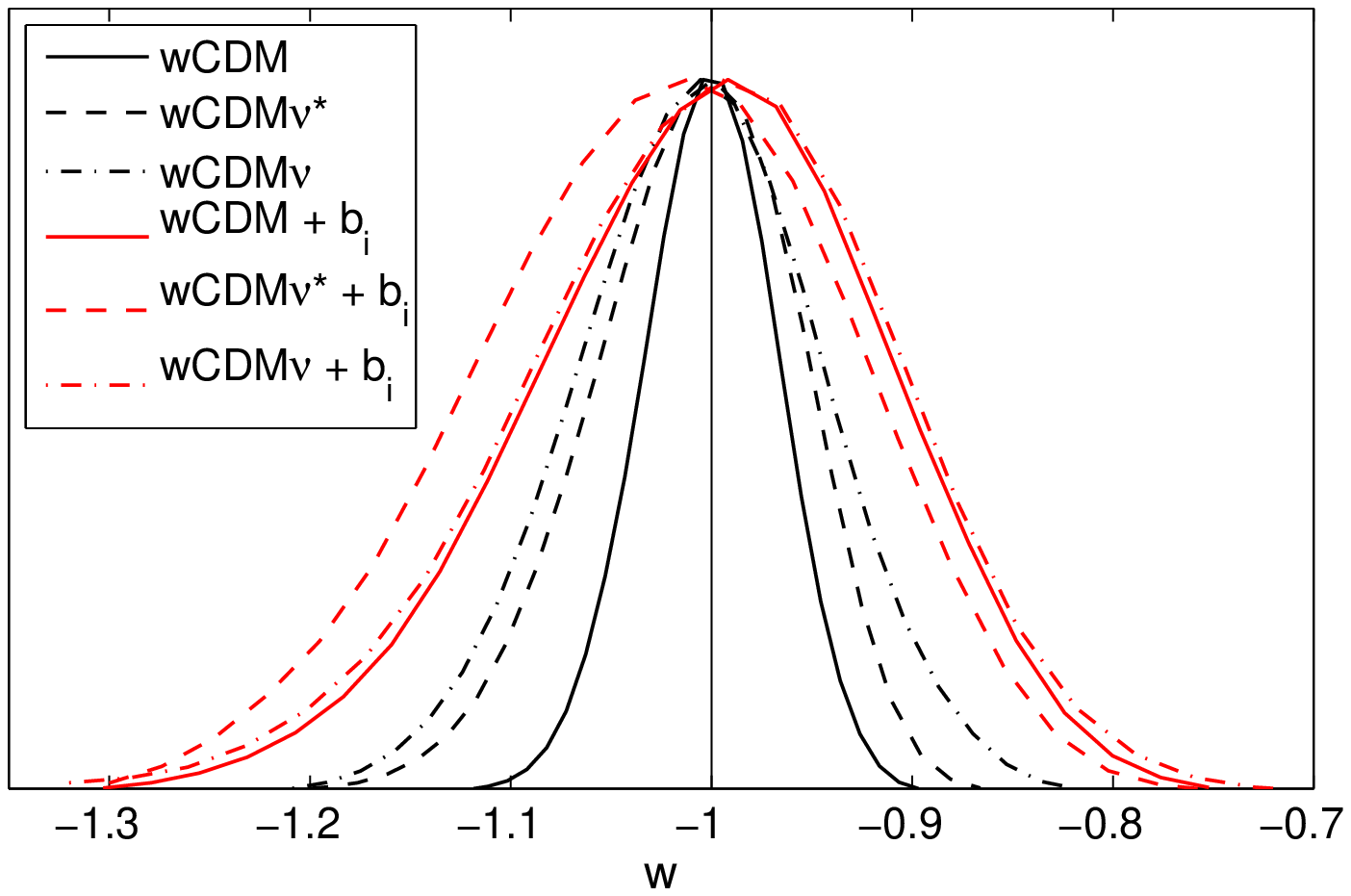}
\includegraphics[width=0.45\textwidth]{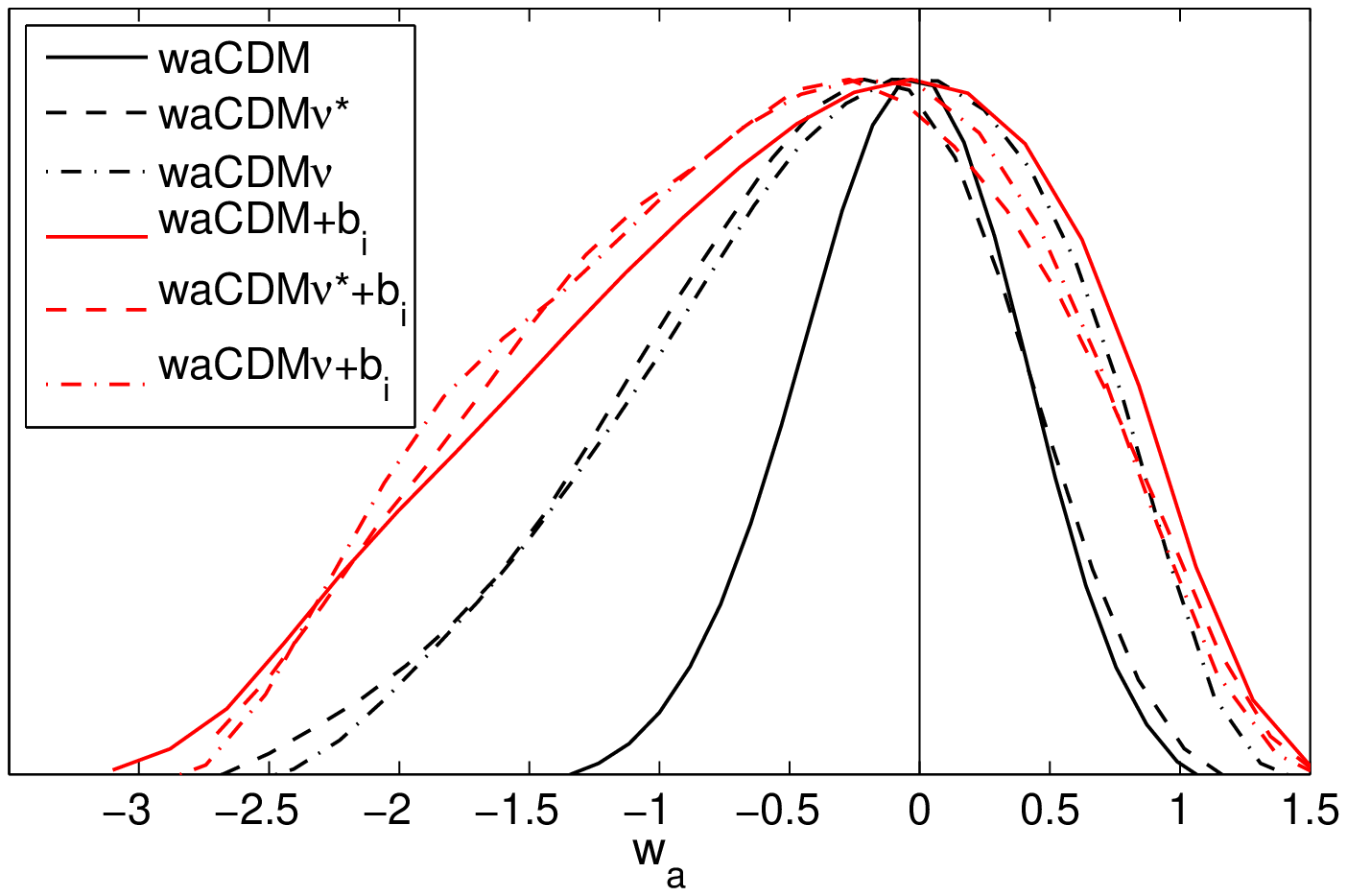}
\caption{\label{fig:figure12} 1D marginalized limits on the equation of state parameters $w$ and $w_a$ for massless and massive neutrino models.
Models marked with a * are for a neutrino mass of $\sum m_{\nu}=0.047$ eV. The constraints on $w$ in the left panel are only from constant $w$ models.
Models with a redshift-evolving bias are plotted in red, whereas models that assume perfect knowledge of bias with $b=1$ are plotted in black.}
\end{figure*}
\begin{table*}[htp]
\scriptsize
\caption{\label{tab:table3a} Recovered marginalized estimates for $w$ and $w_a$ in models with and without massive neutrinos. The neutrino masses used are $\sum m_{\nu}=0.047$ and
$\sum m_{\nu}=0.235$ eV. All results are for Planck and DES synthetic data up to $l_{\text{max}}=300$.}
\begin{ruledtabular}
\begin{tabular}{c c c c c c c}
\rule{0pt}{4ex} Parameter & wCDM &  wCDM$+\text{b}$ & $\text{wCDM}+ \text{b}_{i}$ & waCDM & waCDM$+\text{b}$  & $\text{waCDM}+ \text{b}_{i}$
\\ [0.5ex]
\colrule
\rule{0pt}{2.5ex}$w$   & $ -1.00 \pm 0.03 $ & $ -1.00 \pm 0.054 $ & $-1.00 \pm 0.085 $ & $-0.98 \pm 0.20 $ & $-0.92 \pm 0.30 $ & $ -0.82 \pm 0.36 $  \\
\rule{0pt}{2.5ex}$w_a$   &-&-&-& $-0.06 \pm 0.42 $ & $-0.23 \pm 0.77 $ & $ -0.50 \pm 0.92 $   \\
\hline
\hline
\rule{0pt}{4ex} $\sum m_{\nu}=0.047$ eV & wCDM$\nu$ &  wCDM$\nu+$b & wCDM$\nu+\text{b}_{i}$ & waCDM$\nu$ & waCDM$\nu+$b  & waCDM$\nu+ \text{b}_{i}$ 
\\ [0.5ex]
\colrule
\rule{0pt}{2.5ex}$w$   & $ -1.01 \pm 0.05 $ & $ -1.017 \pm 0.058 $ & $-1.026 \pm 0.088 $ & $-0.83 \pm 0.28 $ & $-0.88 \pm 0.32 $ & $ -0.80 \pm 0.35 $  \\
\rule{0pt}{2.5ex}$w_a$   &-&-&-& $-0.49 \pm 0.72 $ & $-0.37 \pm 0.83 $ & $ -0.58 \pm 0.89 $   \\
\hline
\hline
\rule{0pt}{4ex} $\sum m_{\nu}=0.235$ eV & wCDM$\nu$ &  wCDM$\nu+$b & wCDM$\nu+\text{b}_{i}$ & waCDM$\nu$ & waCDM$\nu+$b  & waCDM$\nu+ \text{b}_{i}$ 
\\ [0.5ex]
\colrule
\rule{0pt}{2.5ex}$w$   & $ -1.00 \pm 0.06 $ & $ -1.00 \pm 0.060 $ & $-1.00 \pm 0.088 $ & $-0.90 \pm 0.29 $ & $-0.93 \pm 0.32 $ & $ -0.80 \pm 0.35$  \\
\rule{0pt}{2.5ex}$w_a$   &-&-&-& $-0.30 \pm 0.75 $ & $-0.22 \pm 0.82 $ & $ -0.57 \pm 0.88$   \\
\end{tabular}
\end{ruledtabular}
\end{table*}
\begin{table*}[htp]
\scriptsize
\caption{\label{tab:table3b} Marginalized upper limits (one-tail 95\% C.L.) on the sum of neutrino masses for $\sum m_{\nu}=0.047$ and $\sum m_{\nu}=0.235$ eV, not including models with a 
time-variable dark energy equation of state (see Table \ref{tab:table3c}). For the higher neutrino mass we show the recovered value of the sum of
neutrino masses with 1$\sigma$ errors. All results are for Planck and DES synthetic data up to $l_{\text{max}}=300$.}
\begin{ruledtabular}
\begin{tabular}{ccccccc}
\rule{0pt}{3ex} $\sum m_{\nu}=0.047$ eV & $\Lambda$CDM$\nu$ &  wCDM$\nu$ & $\Lambda$CDM$\nu+$b & wCDM$\nu+$b & $\Lambda$CDM$\nu+\text{b}_{i}$ & wCDM$\nu+\text{b}_{i}$ 
\\ [0.5ex]
\colrule
%\rule{0pt}{4ex}$\sum m_{\nu} $ & $ 0.235 \pm 0.0198 $ & $ 0.237 \pm 0.034 $ & $ 0.216 \pm 0.1090 $&$ 0.226 \pm 0.11 $ & $ 0.243 \pm 0.076 $ &$ 0.2456 \pm 0.1341 ??? $ \\
\rule{0pt}{4ex} (95\% C.L.) $\bf{\sum m_{\nu}}$ &$<$ 0.08 &$<$ 0.10&$<$ 0.30 &$<$ 0.31 & $<$ 0.18&$<$ 0.19 \\
\rule{0pt}{4ex}$\sum m_{\nu} $ & $ 0.046 \pm 0.02 $ & $ 0.0514 \pm 0.029 $ & $ 0.133 \pm 0.089 $&$ 0.141 \pm 0.094 $ & $ 0.09 \pm 0.054$ &$ 0.098 \pm 0.0563 $ \\
\hline
\hline
\rule{0pt}{3ex} $\sum m_{\nu}=0.235$ eV & $\Lambda$CDM$\nu$ &  wCDM$\nu$ & $\Lambda$CDM$\nu+$b & wCDM$\nu+$b & $\Lambda$CDM$\nu+\text{b}_{i}$ & wCDM$\nu+\text{b}_{i}$ 
\\ [0.5ex] 
\colrule
\rule{0pt}{4ex} (95\% C.L.) $\bf{\sum m_{\nu}}$ &$<$ 0.27 & $<$ 0.29 &$<$ 0.40 &$<$ 0.41 & $<$ 0.36& $<$ 0.36 \\
\rule{0pt}{4ex}$\sum m_{\nu} $ & $ 0.235 \pm 0.02 $ & $ 0.237 \pm 0.034 $ & $ 0.216 \pm 0.11 $&$ 0.226 \pm 0.11 $ & $ 0.243 \pm 0.076 $ &$ 0.239 \pm 0.08 $ \\
%\rule{0pt}{4ex} (95\% C.L.)two tail $\bf{\sum m_{\nu}<}$ &0.2739 & 0.301727 & 0.424749 &0.4393 & 0.3784 & 0.3831 \\
\end{tabular}
\end{ruledtabular}
\end{table*}
\begin{table}[htp]
\scriptsize
\centering
\caption{\label{tab:table3c} Same as Table \ref{tab:table3b} but for models with 
a time-variable dark energy equation of state.}
\begin{ruledtabular}
\begin{tabular}{cccc}
$\sum m_{\nu}=0.047$ eV& waCDM$\nu$ & \dots$+\text{b}$ & \dots $+\text{b}_{i}$ 
\\ [0.5ex]
\colrule
\rule{0pt}{4ex} (95\% C.L.) $\bf{\sum m_{\nu}}$ &$<$ 0.147 &$<$ 0.333 &$<$ 0.203 \\
\rule{0pt}{4ex} $\sum m_{\nu} $ & $ 0.075 \pm 0.042 $ &$ 0.153 \pm 0.099 $ &$ 0.102 \pm 0.06$ \\
\hline
\hline
\rule{0pt}{4ex}  $\sum m_{\nu}=0.235$ eV & waCDM$\nu$ & \dots$+\text{b}$ & \dots $+\text{b}_{i}$ 
\\ [0.5ex]
\colrule
\rule{0pt}{4ex} (95\% C.L.) $\bf{\sum m_{\nu}}$ &$<$ 0.33 &$<$ 0.42 &$<$ 0.38 \\
\rule{0pt}{4ex} $\sum m_{\nu} $ & $ 0.246 \pm 0.06 $ &$ 0.234 \pm 0.116 $ &$ 0.251 \pm 0.08$ \\
\end{tabular}
\end{ruledtabular}
\end{table}

\subsection{Summary}
\label{sec:Discussion}
We summarize our results on the sum of neutrino masses $\sum m_{\nu}$ in Fig. \ref{fig:figure10} and Tables \ref{tab:table3b}-\ref{tab:table3c}
for $\Lambda$CDM$\nu$, wCDM$\nu$ and waCDM$\nu$ models, while allowing for 
uncertainty in galaxy bias. We have shown that the best upper limit (95\% C.L.) for the sum of
neutrino masses from DES+Planck can reach $\Sigma m_\nu < 0.08 $ eV if we assume perfect knowledge of galaxy bias and $w=-1$ (0.10 eV if $w\ne-1$; 0.30 eV if we vary galaxy bias but not $w$). Assuming
that galaxy bias evolves with redshift and allowing $w$ to differ from $-1$ degrades the upper limit on the neutrino mass to 0.19 eV. Finally, we obtain an upper limit of $\Sigma m_\nu < 0.20 $ eV when we allow both $w$ and 
galaxy bias to evolve with redshift.  These upper limits are for models with $\Omega_{\nu} = 0.001$, which corresponds to a
fiducial mass of $\Sigma m_\nu = 0.047$ eV. In Table \ref{tab:table3c} we show our estimates for the upper limits on sum of neutrino masses for the 
above case, as well as the case where $\Omega_{\nu} = 0.005$, and a mass of $\Sigma m_\nu = 0.235$ eV. For the latter, we also provide the recovered values of $\Sigma m_\nu$ and 1$\sigma$ errors. 

For models with $\Sigma m_\nu = 0.235$ eV, where dark energy is not a cosmological constant ($w\ne-1$ or w(a)), the sum of neutrino masses is 
recovered to within $0.2\sigma$. We also find that whatever dark energy model we choose as the fiducial model, adding a single galaxy bias parameter
or using the 7 bias parameter model recovers the sum of neutrino masses to better than $0.2\sigma$. One interesting aspect of this work is that 
the upper limit on $\Sigma m_\nu$ does not increase significantly when we allow $w$ to vary with redshift, once we also include a 
parametrization of galaxy bias.

In Fig. \ref{fig:figure12} and Table~\ref{tab:table3a} we show our constraints on $w$ and $w_{a}$ for models with and without massive neutrinos,
taking into account our two galaxy bias models. Our strongest constraint on $w$ is from the wCDM model, where $w=-1.00 \pm 0.03$. In models with massive neutrinos
the error on $w$ increases by a factor of 2. In wCDM$\nu$ models that include a single bias parameter, the error remains roughly constant
at 5-6\%, although in a model with the minimum neutrino mass $\Sigma m_\nu = 0.047$ eV, the $w$-$\Sigma m_\nu$ degeneracy means that the
recovered value of $w$ is $w=-1.017 \pm 0.058$. 

In fact, the error on $w$ remains roughly the same (5-6\%) once we add a single bias parameter, regardless of whether the fiducial cosmology includes massless or massive neutrinos. Similarly, when we add 7 bias 
parameters to either a wCDM model or a wCDM$\nu$, the errors on $w$ remain roughly the same at 8-9\%. In a model where $\Sigma m_\nu = 0.047$eV,
we obtain the mean value of $w=-1.026 \pm 0.088$.

We find that our estimates of $w$ are unbiased in models with a higher sum of neutrino masses, regardless of the bias model used. In waCDM$\nu$ models,
the addition of massive neutrinos also increases the errors on $w_0$ and $w_a$, but their errors stay roughly constant once we include either a 
single bias model or a 7 parameter bias model. While the most accurate determination of $w_0$ and $w_a$ comes from a waCDM model with massless
neutrinos, the waCDM$\nu$ model with a higher fiducial neutrino mass and a single bias yields constraints that are more accurate than the 
model with bias and a minimal mass of neutrinos. In waCDM$\nu$ models with 7 bias parameters the recovered values of $w_0$ and $w_a$ and their
errors are very similar regardless of the neutrino mass we assume.

%%%%%%%%%%%%%%%%%%%%%%%%%%%%%%%%%%%%%%%%%%%%%%%%%%%%%%%%%%%%%%%%%%%%%%%%%%%%%%%%%%%%%%%%%%%%%

\section{Assumptions}
\label{sec:Assumptions}
There are several difficulties involved with obtaining a measurement or an upper limit on the sum of neutrino masses. In this paper we
have have not considered the effect of non-linear scales on neutrino mass constraints. Such an analysis will require the detailed
understanding of massive neutrino perturbations calibrated against N-body simulations. Furthermore, galaxy bias will no longer 
be scale-independent as was assumed in this work \cite{2014JCAP...03..011V}.

We also do not model the effects of redshift-space distortions, nor do we worry about the accuracy of the Limber and the 
small angle approximations. Instead, we have checked how much the constraints change if we exclude the range in multipole 
space from $l=2$ to $l=30$ from our likelihood analysis. The results are discussed in Appendix \ref{6six_ccc}.

%%%%%%%%%%%%%%%%%%%%%%%%%%%%%%%%%%%%%%%%%%%%%%%%%%%%%%%%%%%%%%%%%%%%%%%%%%%%%%%%%%%%%%%%%%%%%

\section{Conclusion}
\label{sec:Conclusion}

In this work we have carried out a joint constraints analysis of how well the angular clustering of galaxies in photo-z shells in DES, in combination 
with a CMB experiment like Planck, will constrain the sum of neutrino masses and the dark energy equation of state. Our main results are:
\begin{enumerate}
 \item Adding DES galaxy clustering to CMB data reduces the error on $w$ and the sum of neutrino masses $\sum m_{\nu}$ by a factor of 10 compared
 to errors from a Planck only analysis.
 \item DES galaxy clustering in combination with CMB data can place competitive constraints on the sum of neutrino masses
 in the region of 0.1 to 0.2 eV, assuming a minimum mass of $\Sigma m_\nu=0.047$ eV and a perfect knowledge of galaxy bias. 
 \item For the $\Lambda$CDM and wCDM models, the upper limits on the sum of neutrino masses $\sum m_{\nu}$ are 0.08 eV and 0.10 eV (95\% C.L.),
 suggesting that DES could distinguish between the normal and the inverted hierarchy provided galaxy bias is known.
 \item We find that once we include a galaxy bias parametrization, whatever the bias model (constant or redshift-evolving), changing the dark energy equation
  of state parameter does not change constraints on the sum of neutrinos by very much.
 \item We find that a 7 parameter bias model determines $\sum m_{\nu_{}}$ more accurately than a single bias model and excludes the region
 in parameter space, where $\Sigma m_\nu$ is small. 
 \item The smallest error on constant $w$ is $\Delta w= 0.03$ in the wCDM model with massless neutrinos.
 \item The errors on constant $w$ in models which include a single bias parametrization, are 5-6\% and 8-9\%
 in a 7 parameter bias model. This is true regardless of whether the model has massive or massless neutrinos.
 \item These results are robust to assumptions about the galaxy bias models and the dark energy equation of state. We therefore conclude that
 adding more DES probes to this analysis will further improve constraints on both the sum of neutrino masses and $w$.
\end{enumerate}

%%%%%%%%%%%%%%%%%%%%%%%%%%%%%%%%%%%%%%%%%%%%%%%%%%%%%%%%%%%%%%%%%%%%%%%%%%%%%%%%%%%%%%%%%%%%%

\appendix

%%%%%%%%%%%%%%%%%%%%%%%%%%%%%%%%%%%%%%%%%%%%%%%%%%%%%%%%%%%%%%%%%%%%%%%%%%%%%%%%%%%%%%%%%%%%%

\section{Improvement over Planck when adding DES data}
\label{6six_bb}

Before combining our synthetic CMB and DES datasets we carried out an assessment of our Planck constraints compared to other forecasts in the literature.
We find that parameter errors from the analysis of our Planck chains are very similar to those obtained by \cite{2005PhRvD..71h3008L} and
\cite{2010PhRvD..82l3504G}, while our errors are somewhat smaller than those in \cite{2006JCAP...10..013P}. These differences can be attributed
to factors such as the number of channels used, the assumed sensitivity and the sky coverage for Planck. Therefore our Planck errors
are robust, given the various assumptions.

\begin{table}[t]
\caption{\label{tab:table4b} 1$\sigma$ marginalized errors from Planck only runs compared with errors obtained when we add
DES synthetic data up to $l_{\text{max}} = 100$ and $l_{\text{max}} = 300$. Galaxy bias is set to $b=1$.
}
\begin{ruledtabular}
\begin{tabular}{C{2.1cm}C{1.4cm}C{1.95cm}C{1.95cm}}
Parameter &Planck& \head{0.5cm}{\text{Planck+DES} \text{$\lmax  = 100$}}  & \head{0.5cm}{\text{Planck+DES} \text{$\lmax  = 300$}}
\\ [0.5ex]
\colrule
\rule{0pt}{2.5ex} \textbf{\lcdm} \\
$\Omega_{\Lambda}$   &   0.0048  & 0.0046&0.0033\\
$t_{\text{age}}$   &   0.016 & 0.015&0.012\\
$\sigma_{8}$   &   0.0046    & 0.0044 &0.0025\\
$H_{0}$          &   0.42 & 0.41&0.30 \\
\rule{0pt}{2.5ex} \textbf{wCDM} \\ 
$w$             &   0.34  &0.047&0.033 \\
\rule{0pt}{2.5ex} \textbf{waCDM} \\ 
$w_0$             &   0.39  &0.32 & 0.2 \\
$w_{a}$          &  1.0  &0.70 &0.42\\
\textbf{$\Lambda$CDM$\mathbf{\nu}$} \\
\rule{0pt}{2.5ex}$\Omega_{\nu_{}}=0.001$   &      0.005     &0.00064&0.00044 \\
$\Omega_{\nu_{}}=0.005$   &       0.006   & 0.00085&0.00048 \\
$\sum m_{\nu_{}}=0.047$ eV  &  0.20      &0.03 &0.02	\\
$\sum m_{\nu_{}}=0.235$ eV  &     0.21    &0.037 &0.02 \\
\end{tabular}
\end{ruledtabular}
\end{table}

Compared to an analysis based on our synthetic Planck data only (Table \ref{tab:table4b}), we find that the addition of galaxy clustering 
from DES gives significant improvement in parameter constraints especially in models where $w \neq -1$, as well as models with
massive neutrinos. This is to be expected since clustering probes both the expansion history and the growth of structure at late times.
Including DES galaxy clustering data (to $l_{\text{max}}=300$) improves the constraints on neutrino mass and on constant $w$ by roughly
an order of magnitude; the improvement on the time-varying dark energy EOS parameters is roughly a factor of 2. Other cosmological parameters are 
already tightly constrained by the CMB, so the errors do not change appreciably.

As discussed in Section \ref{sec:Results}, throughout this paper we include galaxy clustering up to $l_{\text{max}}=300$ 
in order to remove scales in the non-linear regime of clustering. Since $l=300$ is comparable to the scale where non-linearity becomes 
important at low-redshift, it is worth exploring the impact on dark energy and neutrino mass constraints of making a more conservative choice.
In Table \ref{tab:table4b}, we show results for both the fiducial choice of $l_{\text{max}}=300$ and a more conservative cutoff scale of
$l_{\text{max}}=100$. We see that the latter choice degrades the CMB+DES constraints on both the dark energy EOS parameters and the neutrino mass by $\sim$50\%.
\begin{table*}[htp]
\caption{\label{tab:table4gbb} Percentage increase in marginalized parameter errors when discarding multipoles up to $l=30$. Shown are
models where there are noticeable differences compared with using full multipole information.}
\begin{ruledtabular}
%\begin{tabular}{C{1.0cm}C{1.0cm}C{1.0cm}C{1.0cm}C{1.0cm}C{1.0cm}C{1.0cm}C{1.0cm}}
\begin{tabular}{@{}cccccccc@{}}
\rule{0pt}{2.5ex}Parameter &wCDM&$\Lambda$CDM$\nu^{*}$&wCDM$\nu^{*}$&$\text{wCDM}\nu^{*}+b_i$&$\text{wCDM}\nu$&$\Lambda\text{CDM}\nu+b_i$&$\text{wCDM}\nu+b_i$
%Parameter &wCDM&$\Lambda$CDM$\nu^{*}$&wCDM$\nu^{*}$&$\text{wCDM}\nu^{*}+b_i$&$\text{wCDM}\nu$&$\Lambda\text{CDM}\nu+b_i$&$\text{wCDM}\nu+b_i$
\\ [0.5ex]
\colrule
\rule{0pt}{2.5ex}$H_{0}$ &   5.2\% &8.4\% &  6\%  & 18.2\%&4.4\%&5.7\% &  21.7\% \\
$t_{\text{age}}$   &   5.5\%&10\% &  6\%  & 9.6\%&6\%&7\% & 28\%\\
$\Omega_{\Lambda}$    & 5.2\% &8.3\% &   6\% & 14.7\%&5.4\%&5\% & 22\%\\
$\sigma_{8}$    &   2.7\% & 0\% &   5.4\% & 12.3\%&3.8\%&10\% &   28\%\\
$\sum m_{\nu_{}}$    &  n/a&11.8\% &   4\%&  20.7\%&0\%&9\% &  38\%\\
$w$              &   4\% &n/a &  4.7\%& 17\%&2.8\%& n/a&   19\%\\
\end{tabular}
\end{ruledtabular}
\end{table*}

%%%%%%%%%%%%%%%%%%%%%%%%%%%%%%%%%%%%%%%%%%%%%%%%%%%%%%%%%%%%%%%%%%%%%%%%%%%%%%%%%%%%%%%%%%%%%

\section{Discarding large-scale information due to uncertainties in redshift-space distortion}
\label{6six_ccc}
Since we use the Limber and small angle approximations to compute the galaxy clustering and because at large scales one has to worry about redshift-space
distortions, we have run a number of chains to see the effect of discarding low multipole information up to $l_{\text{min}} = 30$ from our 
likelihood analysis. We present our results in Table \ref{tab:table4gbb},
where we quote the largest error increase for a given parameter in each model. In \lcdm and wCDM models with massless neutrinos, the
accuracy of parameter estimation is not affected, and the parameter errors in \lcdm remain the same. In the wCDM model
the parameters most affected by discarding large-scale data are $H_0$, $\Omega_{\Lambda}$ and $t_{\text{age}}$ with roughly a 5\%
increase in the error bars. The error on the EOS parameter $w$ increases by 4\%. 

In the $\Lambda$CDM$\nu$ model with the minimal neutrino mass of $\sum m_{\nu_{}} = 0.047$ eV, the recovered neutrino mass is biased high by 1.4\%,
while all other parameter estimates and errors remain the same. The upper limit on the sum of neutrino masses is still 
$\Sigma m_\nu < 0.08$ eV (95\% C.L.). Allowing for $w\ne-1$ in the same minimal neutrino mass model biases the estimate for
the neutrino mass by 4\%. The error on $w$ increases by 4.7\%. The errors on $H_0$, $\Omega_{\Lambda}$ and $t_{\text{age}}$ and $\sigma_8$ are
roughly 6\% higher. All other estimates of the fiducial parameters are unaffected.
Once we include a redshift-evolving bias model and $w\ne-1$, the upper limit in the minimal mass model increases to 0.23 eV (95\% C.L.), and $w$ is biased 
by 4.6\%. The error on the sum of neutrino masses is 20\%, and other parameters are determined accurately.

In the $\Lambda$CDM$\nu$ model with the higher neutrino mass, both the estimates of the parameters and their errors are unchanged. The wCDM$\nu$
model with the higher mass also recovers the cosmological parameters with the same accuracy as when one uses all the multipoles. The errors on
$w$ in this model increase by 2.8\%. In the higher neutrino mass model where the galaxy bias evolves with redshift, parameter estimates are
unbiased, and the biggest increase in error is for $\sigma_8$ and $\Sigma m_\nu$ of 10\% and 9\% respectively. The upper limit in the 
redshift-evolving bias model and $w\ne-1$ is 0.42 eV (95\% C.L.), and the $\Sigma m_\nu$ is biased by 1\%. The estimate of $w$ is unbiased, 
and the error on $w$ increases by 19\%. There is a 1\% increase in the estimate of $\sum m_{\nu}$ with an error increase of 38\% from 0.08 to 0.11 eV.

%%%%%%%%%%%%%%%%%%%%%%%%%%%%%%%%%%%%%%%%%%%%%%%%%%%%%%%%%%%%%%%%%%%%%%%%%%%%%%%%%%%%%%%%%%%%%

\section{Uncertainty on Galaxy Bias}
\label{bias_plots}

\begin{figure}[htp]
\includegraphics[width=0.42\textwidth]{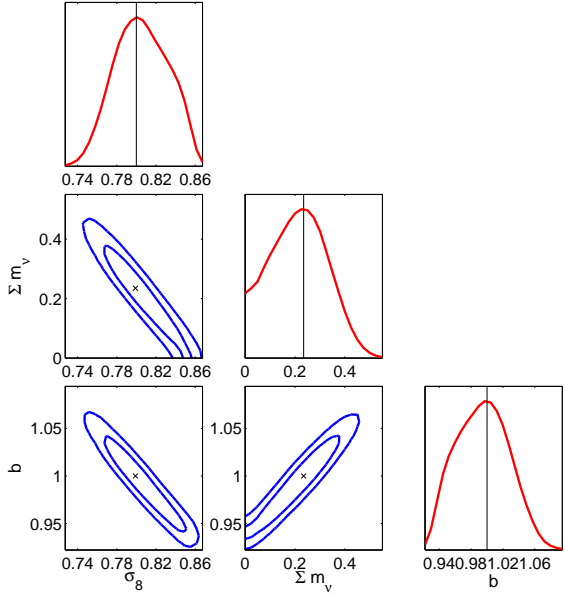}
\caption{\label{fig:figure13} 68 and 95\% likelihood contours and 1D marginalized constraints for $\Lambda$CDM$\nu+b$ with $\Omega_{\nu}=0.005$ including a
single free bias parameter $b$. The input values of the parameters are marked with $\times$.}
\end{figure}

The uncertainty on the value of galaxy bias in \lcdm and wCDM models is less than 0.5\pcnt\, and 1.0\pcnt\, respectively. In models with massive neutrinos only or 
neutrinos and dark energy together, we can recover the true galaxy bias with a precision between 3-5\%. In a model with seven bias parameters,
the error on $b_i$ increases with redshift from 3-5\%. Galaxy bias uncertainty is of the same order in models with seven bias parameters and a 
time-variable dark energy EOS parameter $w(a)$. The errors on galaxy bias are shown in Table \ref{tab:table5a}. Fig. \ref{fig:figure13} shows likelihood contours
\begin{figure}[htp]
\includegraphics[width=0.46\textwidth]{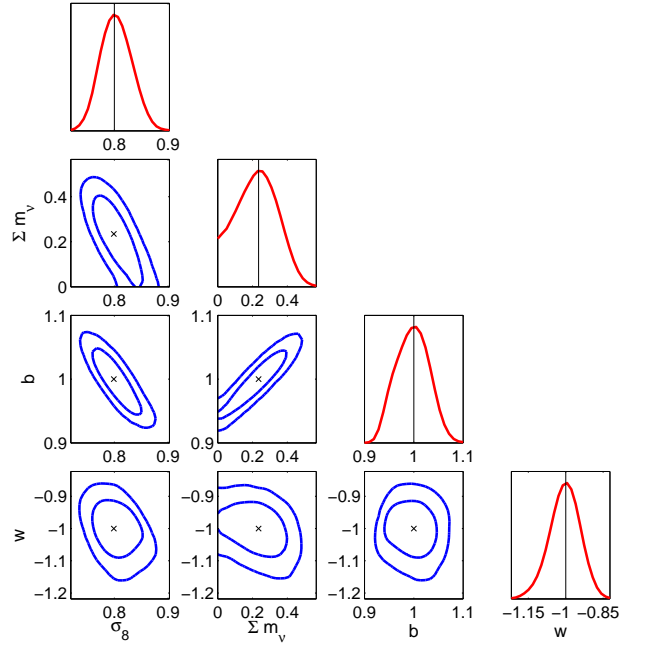}
\caption{\label{fig:figure14bbb} Same as \ref{fig:figure13} but for wCDM$\nu+b$ with $\Omega_{\nu}=0.005$ and $w\ne-1$.}
\end{figure}
and 1D marginalized constraints for a subset of parameters in a $\Lambda$CDM$\nu+b$ model with $\Omega_{\nu}=0.005$. Galaxy bias is highly degenerate with neutrino mass and $\sigma_8$ as expected. In Fig. \ref{fig:figure14bbb} 
we show the constraints in a wCDM$\nu+b$ model with the same fiducial neutrino mass. We find that in a model with neutrinos and $w$, galaxy bias is non-degenerate with $w$.
Fig. \ref{fig:figure15} and Fig. \ref{fig:figure16} show the constraints in a $\Lambda$CDM$\nu+b_i$ and wCDM$\nu+b_i$ models with $\Omega_{\nu}=0.005$
including a redshift-evolving bias and EOS parameter $w$. The degeneracy between galaxy bias and $\sigma_8$ decreases as a function of redshift in the
wCDM$\nu+b_i$ model. The $w-\sum m_{\nu}$ degeneracy is also less severe compared to any wCDM$\nu$ model. Each of the seven bias parameters is
non-degenerate with $w$.

\begin{table}[t]
\scriptsize
\caption{\label{tab:table5a}%
Galaxy bias results in models with $\sum m_{\nu}= 0.235$eV.}
\begin{ruledtabular}
\begin{tabular}{c c c c c}
\rule{0pt}{3ex} Bias & $\Lambda$CDM$\nu+\text{b}$ & wCDM$\nu+\text{b}$ & $\Lambda$CDM$\nu+\text{b}_{i}$ & $\text{wCDM}\nu+\text{b}_{i}$
\\ [0.5ex]
\colrule
\rule{0pt}{3.0ex} $b_{}=1.00$       & $0.994 \pm 0.032$  & $0.997 \pm 0.031$& -& -  \\
\rule{0pt}{2.0ex} $b_{1}=1.45$ &- &- &$1.45\pm 0.034$ &$ 1.45 \pm 0.038$   \\
\rule{0pt}{2.0ex}  $b_{2}=1.60$ &- &- &$1.60\pm 0.037$ & $1.60 \pm 0.039$  \\
\rule{0pt}{2.0ex}  $b_{3}=1.78$ &- &- &$1.78\pm 0.040$ &  $1.78 \pm 0.041$ \\
\rule{0pt}{2.0ex}  $b_{4}=1.97$ &- &- &$1.97\pm 0.043$ & $ 1.98 \pm 0.044$ \\
\rule{0pt}{2.0ex} $b_{5}=2.19$ &- &- &$2.19\pm 0.045$ & $ 2.19 \pm 0.046$ \\
\rule{0pt}{2.0ex} $b_{6}=2.39$ &- &- &$2.39\pm 0.048$ &$ 2.39 \pm 0.048$ \\
\rule{0pt}{2.0ex}  $b_{7}=2.59$ &- &- &$2.59\pm 0.051$ & $ 2.59 \pm 0.051$ \\
\end{tabular}
\end{ruledtabular}
\end{table}
\begin{figure*}[htp]
\includegraphics[width=0.8\textwidth]{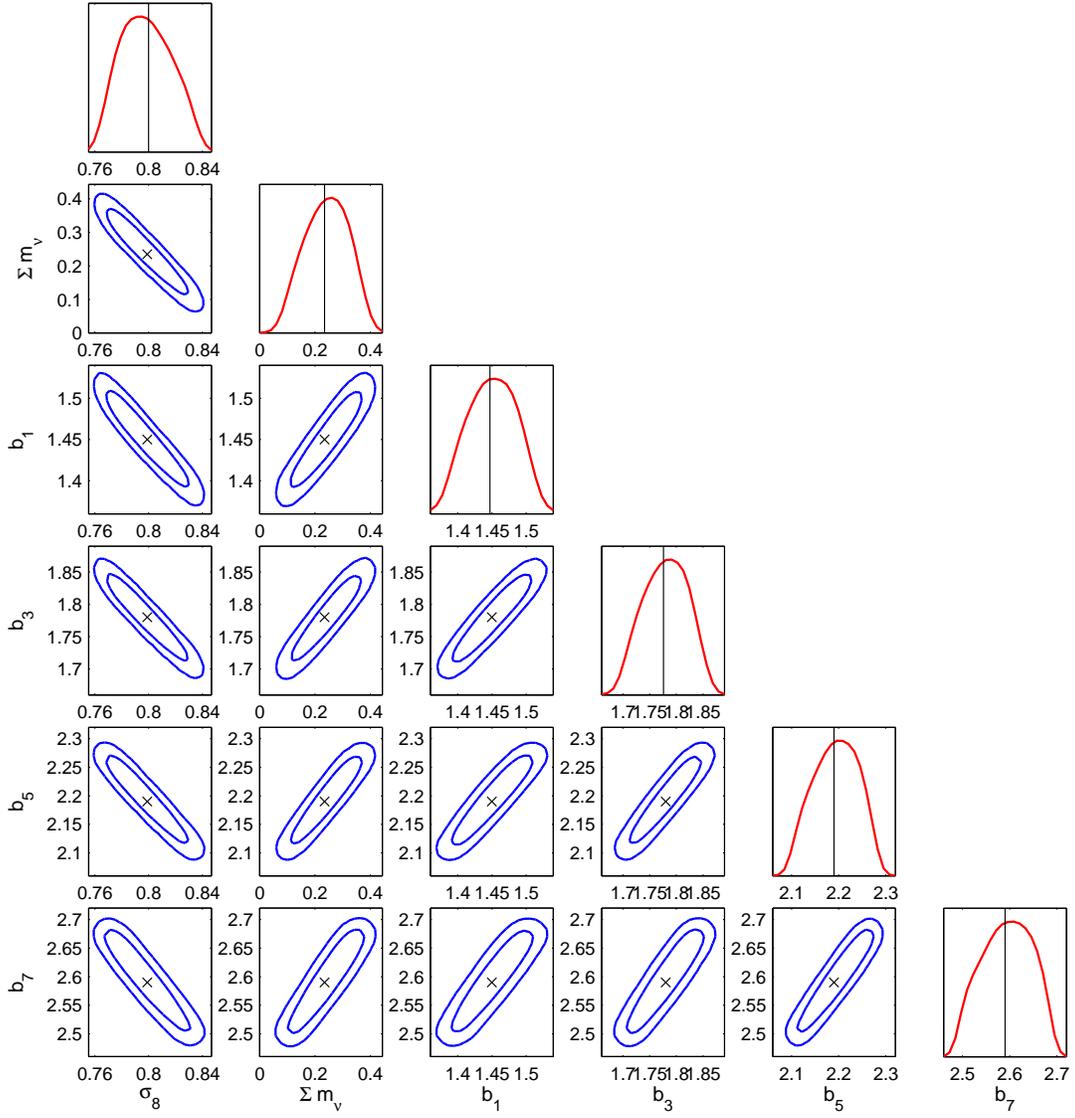}
\caption{\label{fig:figure15} 1D and 2D marginalized constraints for \lcdm$\nu+b_i$ with $\Omega_{\nu}=0.005$ including a redshift-evolving bias.
Shown are degeneracies in a model with 7 bias parameters. We only show 4 of the 7 bias parameters.}
\end{figure*}
\begin{acknowledgments}
A.Z. would like to thank Josh Frieman for his guidance throughout this project. I am grateful to Surhud More, Douglas Rudd, Vinicius Miranda, 
Benedikt Diemer, Douglas Watson, Wayne Hu and Dan Grin for useful discussions. This work was completed in part with resources provided by the
University of Chicago Research Computing Center as well as the Joint Fermilab - KICP Supercomputing Cluster, supported by grants from Fermilab,
Kavli Institute for Cosmological Physics, and the University of Chicago. A.Z. acknowledges support from KICP, the Brinson Foundation, and the U.S. Dept. of 
Energy contract DE-FG02-13ER41958.
\end{acknowledgments}

\newpage
\begin{figure*}[t!]
\includegraphics[width=0.95\textwidth]{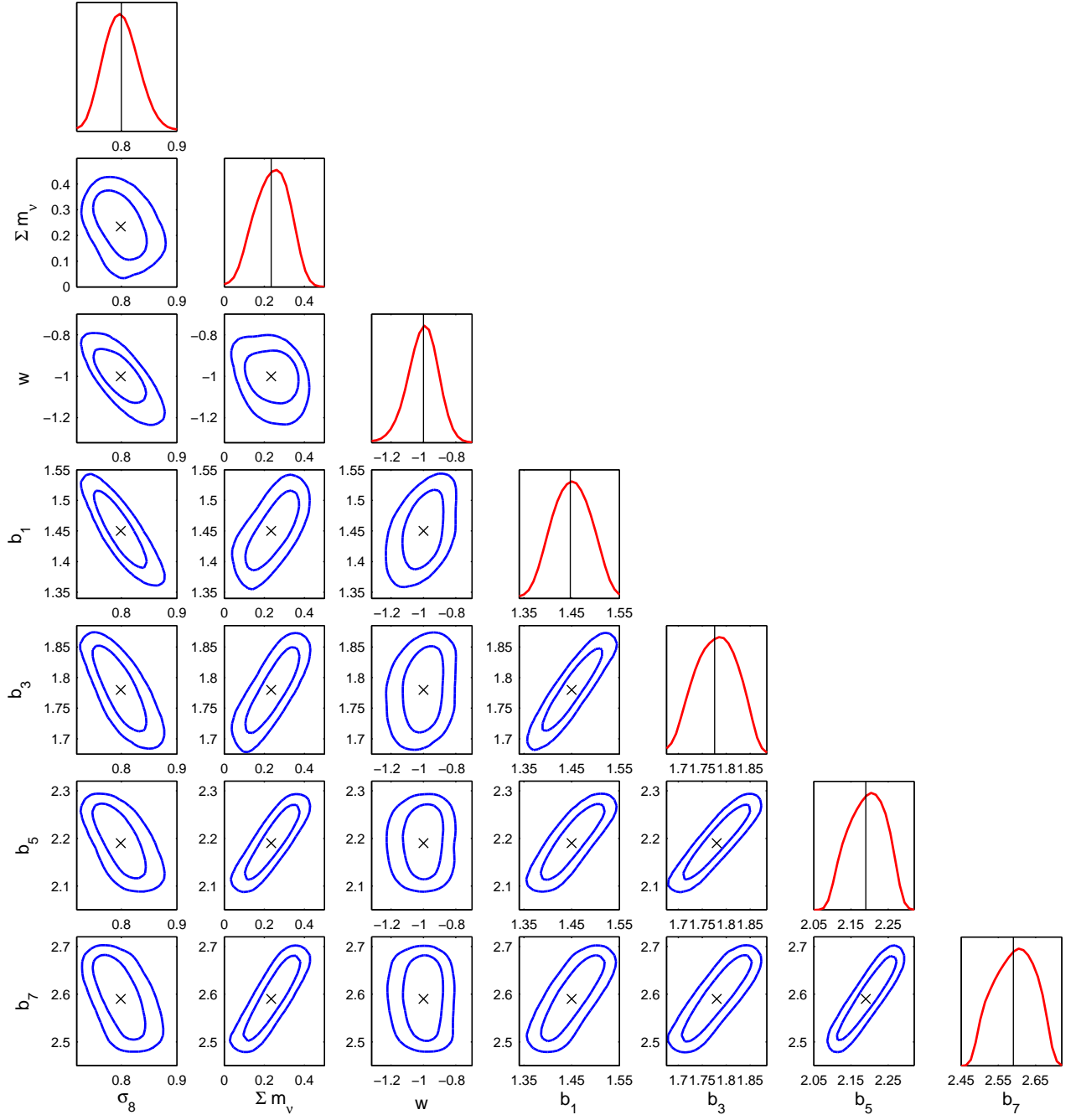}
\caption{\label{fig:figure16} 1D and 2D marginalized constraints for wCDM$\nu+b_i$ with $\Omega_{\nu}=0.005$ including a redshift-evolving bias and
EOS parameter $w$.}
\end{figure*}

\clearpage

%\bibliography{DES_Neutrino_paper_final3}


\end{document}